\documentclass[reprint,amsmath,amssymb,aps,10pt,pre,twocolumn,longbibliography]{revtex4-2}
\usepackage[pdftex]{graphicx}
\usepackage{bm}
\usepackage{bbm}
\usepackage{color}

\newcommand{\revis}[1]{\textcolor{black}{#1}}
\newcommand{\modif}[1]{\textcolor{black}{#1}}

\usepackage{tabularx}

\usepackage{blkarray}
\usepackage{xr}
\newcommand\bigstrutht{\vrule width0pt height 12pt depth0pt\relax}
\newcommand\bigstrutdp{\vrule width0pt height 0pt depth5pt\relax}

\begin{document}

\title{Network analysis for the steady-state thermodynamic uncertainty relation}

\author{Yasuhiro Utsumi}
\affiliation{Department of Electrical and Electronic Engineering, Faculty of Engineering, Mie University, Tsu, 514-8507, Mie, Japan.}

\begin{abstract}
We perform network analysis of a system described by the master equation to estimate the lower bound of the steady-state current noise, starting from the level 2.5 large deviation function and using the graph theory approach.
When the transition rates are uniform, and the system is driven to a non-equilibrium steady state by unidirectional transitions, we derive a noise lower bound, which accounts for fluctuations of sojourn times at all states and is expressed using mesh currents. 
This bound is applied to the uncertainty in the signal-to-noise ratio of the fluctuating computation time of a schematic Brownian computation plus reset process described by a graph containing one cycle.
Unlike the mixed and pseudo-entropy bounds that increase logarithmically with the length of the intended computation path, this bound depends on the number of extraneous predecessors and thus captures the logical irreversibility.
\end{abstract}

\date{\today}

\begin{titlepage}
\maketitle
\end{titlepage}

\newcommand{\mat}[1]{\mbox{\boldmath$#1$}}

\section{Introduction}

The thermodynamic uncertainty relation (TUR) provides a universal trade-off between precision and dissipation~\cite{Horowitz2020}.
In the last decade, the TUR and its relatives—the trade-off relation and speed limits—have been discussed from various perspectives, e.g., Ref.~\cite{ShiraishiBook2023}.
Among these, the TUR has been applied to Bayes nets~\cite{Wolpert2020} and Brownian computation models~\cite{utsumi2022computation}.
Such networks, typically large-scale, possess information processing capabilities, making them intriguing subjects from novel aspects of the thermodynamics of computation~\cite{Bennett1982,wolpert2023stochastic}.
However, the TUR bound is anticipated to be weak for large networks because it is formulated in quantities, such as entropy production and activities, that increase as the system size grows. 
Therefore, tightening the bound is necessary for the practical application of TUR to computation models extended to include dynamics. 

Graph theory is a well-established tool for analyzing electric circuit networks~\cite{Bryant1961, Branin1967,Takahashi1969, RasmussenPRX2021}. 
The network is algebraically treated using the circuit matrices, i.e., the incidence matrix, the cycle (loop) matrix, and the cutset matrix. 
This approach aims to systematically reduce the number of free coordinates in the circuit equations or Lagrangians~\cite{RasmussenPRX2021}.
In the present paper, we perform a network analysis of a directed multigraph describing a Markov chain in the steady state. 
Such a graph can describe the Brownian computation plus reset process~\cite{utsumi2022computation, Utsumi2023}. 

For this purpose, we go back to the level 2.5 large deviation function adopted in early TUR studies~\cite{BertiniAIHP2015,BERTINI2015,Barato2015,Gingrich2016, Pietzonka2016, Polettini2016, Gingrich2017, GingrichPRL2017, Garrahan2017}. 
The level 2.5 large deviation function provides the joint probability distribution of the numbers of jumps at all arcs and the sojourn times at all nodes in the limit of long measurement time. 
It derives a formally exact expression of the probability distribution of steady-state current, thus serving as a solid starting point for the network analysis.
We derive a lower bound of the current noise when considering both bidirectional and unidirectional transition processes with uniform transition rates. 
The importance of network topology has been recognized in the studies of the steady-state TUR~\cite{Pietzonka2016,Polettini2016,Gingrich2017} and the level 2.5 large deviation theory~\cite{BertiniAIHP2015,BERTINI2015} in connection to the steady-state fluctuation theorem~\cite{Andrieux2007}. 
These studies have focused on the universal aspect, while in the present paper, we emphasize practical application to large networks, especially to the Brownian computation model. 

A secondary purpose of the paper is to provide an elementary derivation of the level 2.5 large deviation function based on the full-counting statistics (FCS)~\cite{Bagrets2003,Utsumi2007,FlindtPRB2010}. 
The original derivation is rigorous and intricate~\cite{BertiniAIHP2015,BERTINI2015}. 
\modif{The concept of level 2.5 large deviation function was introduced based on stochastic trajectories~\cite{Maes_2008}.}
We aim for our derivation to be an accessible introduction to this concept for the mesoscopic quantum transport community and to make this paper self-contained simultaneously.

The structure of the paper is as follows: 
In Sec.~\ref{sec:level_2.5_large_deviation_function}, we re-derive the level 2.5 large deviation function using the FCS approach and summarize previously known noise lower bounds. 
This section also introduces notations, which we use for our graph theoretical analysis. 
\modif{The symbols are summarized in Supplemental Material~\cite{LisSym}.}
Section~\ref{sec:Lower_bound_of_current_noise} explains our contributions: 
After introducing the circuit matrices, we derive a lower bound of current noise by duality transformation for systems with uniform transition rates. 
In Sec.~\ref{sec:examples}, we apply our bound to a schematic Brownian computation model.
Section~\ref{sec:conclusion} summarizes our results.

\section{level 2.5 rate function and TUR}
\label{sec:level_2.5_large_deviation_function}

\subsection{FCS approach}

The state transition diagram of a continuous-time Markov chain is a directed multigraph $G_{\rm m}=(V,E_{\rm m})$, 
where the set of nodes $V$ and the set of arcs (directed edges) $E_{\rm m}$ represent states and transitions, respectively. 
We focus on a connected graph with a unique steady state for simplicity. 
The direction of an arc corresponds to the direction of a transition. 
We write the arc $e \in E_{\rm m}$ from the tail node $v_+ \in V$ to the head node $v_- \in V$ as a tuple $e=(v_- \leftarrow v_+)$. 
The boundary operator $\partial^\pm$ maps the arc to the node as $\partial^\pm e = v_\pm$. 
The positive (negative) incidence matrix is defined as, 
$D^{\pm}_{v,e} = \delta_{v,\partial^\pm e}$. 
The incidence matrix is $D=D^+-D^-$. 
We denote the reversed arc of $e$ as $-e=(v_+ \leftarrow v_-)$, which satisfies, 
$D^{\pm}_{v,-e} = D^{\mp}_{v,e}$. 
The master equation is, 
\begin{align}
\dot{n}_v & = - \sum_{e \in E_{\rm m}} D_{v,e} a_e(n_v) \, , \\ 
a_e(n_v) & = \sum_{v \in V} \Gamma_e D^+_{v,e} n_v = \Gamma_e n_{\partial^+ e}\, ,
\end{align}
where $n_v$ is the state probability of node $v \in V$, and $\Gamma_e > 0$ is the transition rate associated to arc $e$. 

We introduce the number of transitions through each arc $e$ in its direction, $W_{e}$, and the sojourn time at each node $v$, $\tau_v$. 
Their joint probability distribution function during the measurement time $\tau$ is,  
\begin{align}
{\mathcal P}_\tau( \{ W_{e} \} , \{ \tau_v \} ) =& \int \prod_{e \in E_{\rm m}} \frac{d \lambda_{e}}{2 \pi} \prod_{v \in V} \frac{d \xi_{v}}{2 \pi} 
{\mathcal Z}_\tau( \{ i \lambda_e \}, \{ i \xi_v \} ) 
\nonumber \\ & \times 
e^{-i \sum_{e \in E_{\rm m}} \lambda_e W_e -i \sum_{v \in V} \xi_v \tau_v }  \, ,
\end{align}
where $\lambda_{e}$ and $\xi_v$ are the counting fields for currents~\cite{Bagrets2003} and dwell times~\cite{Utsumi2007}. 
In the limit of long measurement time $\tau$, 
${\mathcal Z}_\tau( \{ i \lambda_{e} \}, \{ i \xi_v \} ) \approx e^{ \tau \Lambda( \{ i \lambda_{e} \}, \{ i \xi_v \} ) }$, 
where $\Lambda( \{ i \lambda_{e} \}, \{ i \xi_v \} )$ is the eigenvalue of the modified (tilted) transition rate matrix, 
\begin{align}
L_{v,v'} 
= \sum_{e \in E_{\rm m}} \Gamma_e \left( D^-_{v,e} e^{i \lambda_e } D^+_{v',e} - D^+_{v,e} D^+_{v',e} \right) + i \xi_v \delta_{v,v'} \, , 
\label{eqn:m_tran_rate}
\end{align}
\modif{with the maximum real part, which continues to the zero eigenvalue corresponding to the steady state when the counting fields are set to zero.}
Then, within the saddle point approximation, 
$\ln {\mathcal P}_\tau ( \{ W_{e} \} , \{ \tau_v \} ) \approx \tau {\mathcal I} ( \{ w_{e} \} , \{ n_v \} )$, 
where the rate function is, 
\begin{align}
{\mathcal I} ( \{ w_{e} \} , \{ n_v \} ) =& \sup_{ x_{e} , y_v \in {\mathbb R} } \biggl( \Lambda ( \{ x_{e} \}, \{ y_v \} )  \nonumber \\ &  -\sum_{e \in E_{\rm m}} x_{e} w_{e} -\sum_{v \in V} y_v n_v \biggl) \, .
\label{eqn:Omega}
\end{align}
The flux $w_{e} = W_{e}/\tau$ and the node state probability $n_{v} = \tau_{v}/\tau$ satisfy the Kirchhoff current law (KCL) and the normalization condition, respectively: 
\begin{align}
\sum_{e \in E_{\rm m}} D_{v,e} w_e =& 0 \, , \label{eqn:KCL_g} \\
\sum_{v \in V} n_v =& 1 \, , \;\;\; (0 \leq n_v \leq 1 ). \label{eqn:norm_1} 
\end{align}
%
We introduce the orthonormalized right and left eigenvectors associated with the eigenvalue $\Lambda$ as,
\begin{align}
\sum_{v'\in V} L_{v,v'} u^R_{v'} = \Lambda u^R_v \, , \;\;\;
\sum_{v\in V} u^L_{v} L_{v,v'}  = u^L_{v'} \Lambda \, . 
\end{align}
By noticing that $u_v^{L}$ and $u_v^{R}$ are orthonormalized and are the functions of $x_e$ and $y_v$, the change of eigenvalue induced by small variations $\delta x_e$ and $\delta y_v$ is calculated as, 
$\delta \Lambda =\sum_{v,v'\in V} u^L_{v} \delta L_{v,v'} u^R_{v'} = \sum_{e \in E_{\rm m}} \delta x_e \Gamma_e e^{x_e } u^L_{\partial^- e} u^R_{\partial^+ e} + \sum_{v \in V} \delta y_v u^L_{v} u^R_{v}$. 
By using this we find that the maximum in Eq.~(\ref{eqn:Omega}) is achieved when $x_e$ and $y_v$ implicitly fulfill, 
\begin{align}
n_v = u^L_{v} u^R_{v} \, , \;\;\;\;
x_e = \ln \frac{w_e}{a_e(n_v)} + \sum_{v \in V} D_{v,e} \ln u^L_{v} \, . \label{eqn:saddle_point}
\end{align}
By substituting these solutions into Eq.~(\ref{eqn:Omega}) and using the KCL (\ref{eqn:KCL_g}), we obtain the level 2.5 large deviation function \modif{(Appendix~\ref{app:derivation_rate_function})}: 
\begin{align}
{\mathcal I}(\{ w_e \}, \{ n_v \}) =& \sum_{e \in E_{\rm m}} a_e(n_v) \psi\left( \frac{w_e}{a_e(n_v)}\right) \, , \label{eqn:rate_func_1} \\
\psi(x) =& x-1-x \ln x \, .
\end{align}

\subsection{Bidirectional and unidirectional processes}

The set of arcs is partitioned into mutually disjoint sets of arcs for unidirectional transitions $E_{\rm uni} = \left \{ e \middle | e \in E_{\rm m} \wedge -e \notin E_{\rm m}  \right \}$ and bidirectional transitions $E_{\rm bi} = \left \{ e \middle | e \in E_{\rm m}  \wedge -e \in E_{\rm m} \right \}$ as $E_{\rm m}=E_{\rm uni} \cup E_{\rm bi}$ and $E_{\rm uni} \cap E_{\rm bi} = \emptyset$. 
The set for bidirectional transitions is further partitioned into the sets for forward transitions $E_{\rm b}$ and backward transitions $\overline{E_{\rm b}}$ as 
$E_{\rm bi} = E_{\rm b} \cup \overline{E_{\rm b}}$
and $E_{\rm b} \cap \overline{E_{\rm b}} = \emptyset$. 
Here and hereafter, we use the overline to represent the set of reversed arcs, 
$\overline{A} = \left \{ -e \middle | e \in A \right \}$. 
In the following, we limit ourselves to the case that there exists $E_{\rm b}$ so that an oriented graph $G=(V,E= E_{\rm uni} \cup E_{\rm b})$ contains a directed rooted spanning tree, $T=(V(T)=V,E(T))$. 
For $e \in E_{\rm b}$, we introduce anti-symmetrized and symmetrized fluxes, 
$j_e = w_e - w_{-e}$ and $g_e = w_e + w_{-e}$, 
and integrate out the latter, 
\begin{align}
\sum_{e \in E_{\rm b}} \sup_{g_e \in \mathbb{R}} \left( a_e \psi\left( \frac{g_e+j_e}{2 a_e}\right) + a_{-e} \, \psi \left( \frac{g_{e}-j_e}{2 a_{-e}}\right) \right)\, . 
\end{align}
This can be done (see e.g. Ref.~\cite{Gingrich2017}) and the result is, 
\begin{align}
{\mathcal I}(\{ j_e\}, \{ n_v \}) =& \sum_{e \in E_{\rm b}} 2 \psi_{\rm bi} \left( \frac{j_e}{2 \sqrt{a_e a_{-e}} } , \frac{{j}_e(n_v)}{2 \sqrt{a_e a_{-e}} }  \right) 
\nonumber \\
& \times \sqrt{a_e a_{-e}} + \sum_{e \in E_{\rm uni}} a_e \, \psi\left( \frac{j_e}{a_e}\right) \, , 
\label{eqn:level25ratefun}
\\
\psi_{\rm bi}(x,y) =&  \sqrt{ 1 + x^2 }  - \sqrt{ 1 + y^2 } \nonumber \\ &- x \left( \sinh^{-1} x -  \sinh^{-1} y  \right) \, . \label{eqn:psi_bi}
\end{align}
Here we write $j_e=w_e>0$ for $e \in E_{\rm uni}$ and introduced, 
$j_e(n_v) = a_e(n_v) - a_{-e}(n_v)$. 
The KCL (\ref{eqn:KCL_g}) for the edge current becomes, 
\begin{align}
\sum_{e \in E} D_{v,e} j_e = 0 \, . \label{eqn:KCL}
\end{align}
The steady-state edge current $j_e(n_v^{\rm st})$, where $n_v^{\rm st}$ is the node state probability in the steady-state, satisfies the KCL. 
The rate function (\ref{eqn:level25ratefun}) takes the maximum at $j_e = j_e(n_v^{\rm st})$ and $n_v=n_v^{\rm st}$ as ${\mathcal I}(\{ j_e(n_v^{\rm st}) \}, \{ n_v^{\rm st} \}) = 0$. 

The node state probabilities in the steady state can be calculated using Kirchhoff-Hill theorem~\cite{Hill1966, Schnakenberg1976, Weidlich1978}: 
\begin{align}
n_v^{\rm st} = \frac{1}{Z} \sum_{\mu=1}^{M_v} \prod_{e \in E( T_{v}^\mu(G) )} \Gamma_e \, . 
\label{eqn:KH}
\end{align}
Here $T_v^\mu(G)$ ($\mu=1,\cdots,M_v$) is the $v$-directed spanning tree and $Z$ is the normalization constant. 
$M_v$ is the number of distinct $v$-directed spanning trees.

\subsection{Mixed and pseudo-entropy bounds}

We are interested in the probability distribution of the weighted sum of the edge currents, 
\begin{align}
w = \sum_{e \in E} d_e j_e \, , 
\;\;\;\; 
( d_e \in \mathbb{R} ) \, , \label{eqn:d} 
\end{align}
\modif{in the limit of long measurement time $\tau$, $\ln {\mathcal P}_\tau ( w ) \approx \tau {\mathcal I}(w)$}, which is obtained by contraction, 
\modif{
\begin{align}
{\mathcal I}(w) = \sup_{ j_e \in \mathbb{R}, n_v \in [0,1] } {\mathcal I} ( \{j_e\}, \{ n_v \}) \, , \label{eqn:pdf_w}
\end{align}
}
subjected to the constraints (\ref{eqn:norm_1}), (\ref{eqn:KCL}) and (\ref{eqn:d}). 
The rate function ${\mathcal I}(w)$ takes the maximum when $w$ equals the average $\langle \! \langle w \rangle \! \rangle = \sum_{e \in E} d_e j_e(n_v^{\rm st})$. 

We introduce a parameter representing the deviation from the average, $\epsilon=w/\langle \! \langle w \rangle \! \rangle -1$. 
Then by substituting $n_v=n_v^{\rm st}$ and $j_e = j_e(n_v^{\rm st})(\epsilon+1)$ to Eq.~(\ref{eqn:pdf_w}) and by using the inequalities 
$\psi_{\rm bi}(x,y) \geq  - (x-y)^2 \left( \sinh^{-1} y \right) /(2 y)$~\cite{Gingrich2016}, 
and (Appendix~\ref{app:nakajima_ineq}), 
\begin{align}
\psi(x) \geq - \frac{(x-1)^2}{2} + \frac{(x-1)^3}{6}-\frac{(x-1)^4}{3}, 
\label{eqn:nakajima_ineq}
\end{align}
we obtain the mixed bound~\cite{Pal2021,Shiraishi2021}: 
\begin{align}
{\mathcal I}(w) \geq& - \modif{ \frac{\epsilon^2}{4} \sigma_{\rm mix} } + \sum_{e \in E_{\rm uni}} a_e(n_v^{\rm st}) \psi^{(3,4)}( w/\langle \! \langle w \rangle \! \rangle ) \, , \label{eqn:mix_all}
\\
\modif{
\sigma_{\rm mix}} =& \sum_{e \in E_{\rm b}} j_e(n_v^{\rm st}) \ln \frac{ a_{e} (n_v^{\rm st}) }{ a_{-e} (n_v^{\rm st}) } + 2 \sum_{e \in E_{\rm uni}} a_{e}(n_v^{\rm st})  \, , \label{eqn:mix}
\end{align}
where $\psi^{(3,4)}$ is the cubic and quartic terms of the right-hand side of Eq.~(\ref{eqn:nakajima_ineq}). 
\modif{Inequality~(\ref{eqn:mix_all}) is a global bound applicable to the full spectrum of current fluctuations.}

In the following, we focus on the \modif{scaled} second cumulant and utilize the edge current vector and the state probability vector for concise presentations. 
We refer to $T^*$ as the cotree of $T$, which contains arcs not in $T$, 
$E(T) \cup E(T^*)=E$ and $E(T) \cap E(T^*) = \emptyset$. 
Hereafter, the number of elements of a set $A$ is denoted as $|A|$. 
The edge current vector 
\modif{
${\bm j}^T = \begin{bmatrix} {\bm j}_t^T & {\bm j}_{c}^T \end{bmatrix} \in \mathbb{R}^{|E|}$} is an $|E|$ component real vector. 
Here 
${\bm j}_t = \begin{bmatrix} j_{t_1}, \cdots, j_{t_{|E(T)|}} \end{bmatrix}^T$
and
${\bm j}_c = \begin{bmatrix} j_{c_1}, \cdots, j_{c_{|E(T^*)|}} \end{bmatrix}^T$
are defined on twigs (arcs of the directed rooted spanning tree $T$) $t_1,\cdots, t_{|E(T)|} \in E(T)$ and on chords (arcs of the cotree $T^*$) $c_1,\cdots, c_{|E(T^*)|} \in E(T^*)$, respectively. 
The state probability vector ${\bm n} \in \mathbb{R}^{|V|}$ is 
${\bm n} = \begin{bmatrix} n_{v_1}, \cdots, n_{v_{|V|}} \end{bmatrix}^T$, 
where, $v_1, \cdots, v_{|V|} \in V$. 

A small deviation $\epsilon \ll 1$ shifts the maximizing parameters as, 
\begin{align}
{\bm j} =& {\bm j}( {\bm n}_v^{\rm st}) + {\bm j}^{\perp} \epsilon +  {\bm j}^{(2)} \epsilon^2/2 + \cdots \, , 
\\
{\bm n} =& {\bm n}^{\rm st} + {\bm \phi} \epsilon +  {\bm n}^{(2)} \epsilon^2/2 + \cdots \, .
\end{align}
By substituting them into Eq.~(\ref{eqn:pdf_w}) and expanding up to second order in $\epsilon$, we obtain, 
${\mathcal I}(w) \approx  - \epsilon^2 \langle \! \langle w \rangle \! \rangle^2 / ( 2 \langle \! \langle w^2 \rangle \! \rangle)$, 
\modif{where $\langle \! \langle w^2 \rangle \! \rangle$ represents the scaled second cumulant expressed as}~\cite{Barato2018},  
\begin{align}
\frac{\langle \! \langle w \rangle \! \rangle^2}{\langle \! \langle w^2 \rangle \! \rangle} = \inf_{ {\bm j}^\perp \in J^{(1)}, \, {\bm \phi} \in P^{(1)} } 
\left( {\bm j}^{\perp} - {\bm j}\left( {\bm \phi} \right) \right)^T {\bm G}^{-1} 
\left( {\bm j}^{\perp} - {\bm j}\left( {\bm \phi} \right) \right) \, . \label{eqn:Sigma_w}
\end{align}
The inverse of diagonal weight matrix is ${\bm G} ={\rm diag} \,{\bm g}({\bm n}^{\rm st})$, where $g_e(n_v) = a_e(n_v) + a_{-e}(n_v)$. 
Here and hereafter, we set $\Gamma_{-e}=0$ for $e \in E_{\rm uni}$. 
The constraints (\ref{eqn:norm_1}), (\ref{eqn:KCL}) and (\ref{eqn:d}) are, 
\begin{align}
J^{(1)} =& \left \{ {\bm j}^\perp \in {\mathbb R}^{|E|} \middle |  \left( \langle \! \langle w \rangle \! \rangle = {\bm d}^T {\bm j}^{\perp}  \right) \wedge 
\left(  {\bm j}^{\perp} \in \ker {\bm D} \right) \right \}  \, , \label{eqn:set_jw1}
\\
P^{(1)}=& \left \{ {\bm \phi} \in {\mathbb R}^{|V|} \middle | {\bm 1}^T {\bm \phi} = 0 \right \}  \, , \label{eqn:set_p1}
\end{align}
where ${\bm d} \in \mathbb{R}^{|E|}$, ${\bm D} \in \mathbb{R}^{|V| \times |E|}$ and ${\bm 1}$ is a real vector whose entries are ones. 

By setting ${\bm \phi}={\bm 0}$ and ${\bm j}^\perp = {\bm j}({\bm n}^{\rm st})$, 
Eq.~(\ref{eqn:Sigma_w}) naturally leads to the pseudo-entropy bound~\cite{Shiraishi2021}: 
\begin{align}
\frac{\langle \! \langle w \rangle \! \rangle^2}{\langle \! \langle w^2 \rangle \! \rangle} \leq  \modif{ \frac{\sigma_{\rm pseudo}}{2} } = \sum_{e \in E} \frac{\left( j_e(n_v^{\rm st}) \right)^2}{ \Gamma_e n_{\partial^+ e}^{\rm st} + \Gamma_{-e} n_{\partial^- e}^{\rm st}  } \, . \label{eqn:pseudo}
\end{align}

Equation~(\ref{eqn:Sigma_w}) is a quadratic optimization problem in $|E|+|V|$ parameters subjected to the constraints~(\ref{eqn:set_jw1}) and (\ref{eqn:set_p1}). 
On the other hand, Eqs.~(\ref{eqn:mix}) and (\ref{eqn:pseudo}) imply that the bound depends on the length of cycles. 
In the following section, we will systematically reduce the number of free parameters. 

\section{Lower bound of current noise}
\label{sec:Lower_bound_of_current_noise}

\subsection{Circuit matrices}

We summarize circuit matrices relevant to our network analysis (see, e.g., Refs.~\cite{Bryant1961, Branin1967} and Appendices of Ref.~\cite{RasmussenPRX2021}). 
We write a fundamental cycle $C_{e_1}$ of length $\ell$ as a tuple, a sequence of one chord $e_1 \in E(T^*)$ and $\ell-1$ twigs $e_{2}, e_{3}, \cdots, e_{\ell} \in E(T)$ or reversed ones $e_{2}, e_{3}, \cdots, e_{\ell}  \in \overline{E(T)}$:
\begin{align}
C_{e_1} =(e_{\ell}, \cdots,e_3,e_2,e_1) \, .
\end{align} 
The head and tail of adjacent arcs, $e_n$ and $e_{n+1}$, ($n=1,\cdots,\ell-1$) share the same node $\partial^- e_{n} = \partial^+ e_{n+1}$. 
The first and last arcs satisfy the periodic boundary condition $\partial^- e_\ell= \partial^+ e_1$.
The fundamental cycle matrix 
${\bm B} \in \mathbb{R}^{|E(T^*)| \times |E|}$ 
indicates which arcs are included in each of $|E(T^*)|$ fundamental cycles: 
\begin{align}
{\bm B} = \begin{bmatrix} -{\bm F}^T & {\bm I}_{|E(T^*)|} \end{bmatrix} \, ,
\end{align} 
where ${\bm I}_{|E(T^*)|}$  is a $|E(T^*)| \times |E(T^*)|$ unit matrix and ${\bm F} \in \mathbb{R}^{|E(T)| \times |E(T^*)|}$,   
\begin{align}
\left( {\bm F} \right)_{t,c} = \mathbbm{1}_{C_c}(-t) - \mathbbm{1}_{C_c}(t) \,, 
\end{align} 
for $c \in E(T^*)$ and $t \in E(T)$. 
The indicator function $\mathbbm{1}_{A}(a)$ equals $1$ if $a \in A$ and equals $0$ if $a \notin A$. 

There is a unique directed path from the root $v_0$ to a node $v$ along the directed rooted spanning tree $T$, which we write as a sequence of twigs $e_1,e_2,\cdots e_\ell \in E(T)$ as, 
\begin{align}
P_{v \leftarrow v_0} =(e_\ell, \cdots , e_2,e_1) \, .
\label{eqn:path}
\end{align} 
Here $\ell$ is the length of the path. 
Similar to the cycle, the head and tail of adjacent arcs share the same node. 
The two endpoints are $\partial^+ e_1=v_0$ and $\partial^- e_\ell=v$. 
We introduce a root-to-node path matrix ${\bm S} \in \mathbb{R}^{|V| \times |E(T)|}$ as a variant of the node-to-datum path matrix~\cite{Bryant1961,Branin1967,Takahashi1969}, 
\begin{align}
\left( {\bm S} \right)_{v,t} = \mathbbm{1}_{P_{v \leftarrow v_0}}(t) \, , 
\label{eqn:pathntegratormatrix}
\end{align} 
which is 1(0) if a twig $t \in E(T)$ is in (not in) the path from the root $v_0$ to the node $v$. 
The fundamental cutset matrix ${\bm Q} \in \mathbb{R}^{|E(T)| \times |E|}$ is then introduced as (Appendix \ref{app:cutsetmatrix}), 
\begin{align}
-{\bm S}^T {\bm D}  = {\bm Q} =  \begin{bmatrix} {\bm I}_{|E(T)|} & {\bm F} \end{bmatrix} \, , \label{eqn:cutsetmatrix}
\end{align} 
which implies that ${\bm S}$ acts as the line integral along the directed rooted spanning tree. 
The fundamental cutset is a minimal set of arcs consisting of one twig and zero or more chords that form the boundary between two regions. 
Its $(t,e)$ component is $1$($-1$) if an arc $e$ is in the fundamental cutset with respect to the twig $t$ and bridges the two regions in the same (opposite) direction of the twig $t$.

The incidence matrix and the cycle matrix satisfy (Appendix \ref{app:cutsetmatrix}), 
\begin{align}
{\bm D} {\bm B}^T =& {\bm 0}_{|V| \times |E(T^*)|} \, , \label{eqn:DBT0} \\
{\bm B} {\bm D}^T =& {\bm 0}_{|E(T^*)| \times |V|} \, , \label{eqn:BDT0}
\end{align} 
where, for example, ${\bm 0}_{|V| \times |E(T^*)|}$ is a $|V| \times |E(T^*)|$ zero matrix.  
The cutset matrix satisfies similar relations: 
\begin{align}
{\bm Q} {\bm B}^T =& {\bm 0}_{|E(T)| \times |E(T^*)|} \, , \label{eqn:QBT0} \\
{\bm B} {\bm Q}^T =& {\bm 0}_{|E(T^*)| \times |E(T)|} \, . \label{eqn:BQT0} 
\end{align} 
Equalities (\ref{eqn:DBT0}), (\ref{eqn:BDT0}), (\ref{eqn:QBT0}) and (\ref{eqn:BQT0}) indicates,  
\begin{align}
\mathrm{im} {\bm B}^T \subseteq & \ker {\bm D} , \, \ker {\bm Q}  \, , \label{eqn:imBTinKerD} \\
\mathrm{im} {\bm D}^T , \, \mathrm{im} {\bm Q}^T   \subseteq & \ker {\bm B}  \, . \label{eqn:imDTinKerB}
\end{align} 
The first (second) inclusion relation corresponds to that between groups of (co)boundaries and (co)cycles~\cite{GrossBook2004}. 
Intuitively, Eqs.~(\ref{eqn:BDT0}) and (\ref{eqn:BQT0}) correspond to $\mathrm{curl} \, (\mathrm{grad}) =0$ in the vector analysis.

\subsection{Duality transformation}

In the following, we limit ourselves to the uniform transition rates, 
\begin{align}
\Gamma_e=\Gamma=1 \, , \;\;\; (e \in E_{\rm m}) \, . \label{eqn:uniform}
\end{align}
In this case, the unidirectional transitions drive the system out of equilibrium. 
In Eq.~(\ref{eqn:Sigma_w}), ${\bm j}\left( {\bm \phi} \right)$  can be separated into the `curl-free' component ${\bm D}^T {\bm \phi} \in {\rm im} {\bm D}^T$ and the source term coming from the unidirectional transitions: 
\begin{align}
{\bm j}\left( {\bm \phi} \right)= {\bm D}^T {\bm \phi} + {\bm \Pi}_{E_{\rm uni}} {\bm D}^{- \, T} {\bm \phi} \, , \;\;\;\; {\bm \Pi}_{E_{\rm uni}} = \sum_{e \in E_{\rm uni}} {\bm e}_e {\bm e}_e^T
\, . \label{eqn:j_par_sou}
\end{align}
where ${\bm e}_e \in \mathbb{R}^{|E|}$ is a unit vector, 
$\left( {\bm e}_e \right)_{e'} = \delta_{e,e'}$. 
One can check 
${\bm \phi} = \left( {\bm 1} {\bm 1}^T /|V| -  {\bm I}_{|V|} \right) {\bm S} {\bm \nu}$, 
where ${\bm \nu} \in \mathbb{R}^{|E(T)|}$, satisfies the constraint (\ref{eqn:set_p1}). 
By substituting it into Eq.~(\ref{eqn:j_par_sou}), we obtain, 
\begin{align}
{\bm j}({\bm \phi}) =& {\bm \Phi}^{-1} {\bm Q}^T {\bm \nu} \, , 
\\
{\bm \Phi}^{-1} =& {\bm I}_{|E|} + \sum_{e' \in E_{\rm uni}} \begin{bmatrix} {\bm e}_{e'} {\bm \Omega}_{e'}^T & {\bm 0}_{|E| \times |E(T^*)|} \end{bmatrix} \, , \label{eqn:Phi_matrix}
\end{align} 
where ${\bm \Omega}_{e} \in \mathbb{R}^{|E(T)|}$ is,  
\begin{align}
\left( {\bm \Omega}_{e} \right)_{t} =& |V(T_{\partial^- t})|/|V| - \mathbbm{1}_{ P_{\partial^- e \leftarrow v_0} } (t) \, .
\end{align} 
Here, we write $T_{\partial^- t}$ as a subtree rooted at $\partial^- t$ obtained by cutting the directed rooted spanning tree $T$ by removing the arc $t$. 
The number of nodes in this subtree is $|V(T_{\partial^- t})| = \sum_{v \in V} \left({\bm S}\right)_{v,t}$. 
For explicit form of ${\bm \Phi}$, see Appendix~\ref{app:Phi}. 

We perform the duality transformation~\cite{Nagaosa1999}, which is in the present context, integrating out the `curl-free' component ${\bm Q}^T {\bm \nu}$ by introducing the auxiliary field ${\bm J} \in \mathbb{R}^{|E|}$ to Eq.~(\ref{eqn:Sigma_w}) as, 
\begin{widetext}
\begin{align}
\frac{\langle \! \langle w \rangle \! \rangle^2}{\langle \! \langle w^2 \rangle \! \rangle}  &= \inf_{ {\bm j}^{\perp} \in J^{(1)}(w), {\bm \nu} \in \mathbb{R}^{|E(T)|} } 
\sup_{ {\bm J} \in \mathbb{R}^{|E|}} 
\left( - \left( {\bm J} - {\bm G}^{-1} {\bm i} \right)^T {\bm G} \left( {\bm J} - {\bm G}^{-1} {\bm i} \right) + {\bm i}^T {\bm G}^{-1} {\bm i} \right)  \\
&= \inf_{ {\bm j}^{\perp} \in J^{(1)}(w), {\bm \nu} \in \mathbb{R}^{|E(T)|} } \sup_{ {\bm J} \in \mathbb{R}^{|E|}} 
\left( - {\bm J}^T {\bm G} {\bm J}  + 2 {\bm J}^T {\bm j}^\perp 
- 2 {\bm J}^T {\bm \Phi}^{-1} {\bm Q}^T {\bm \nu} \right) \, , \label{eqn:dual}
\end{align}
\end{widetext}
where ${\bm i} = {\bm j}^{\perp} - {\bm j}\left( {\bm \phi} \right)$. 
Equation (\ref{eqn:dual}) is equivalent to, 
\begin{align}
\frac{\langle \! \langle w \rangle \! \rangle^2}{\langle \! \langle w^2 \rangle \! \rangle}  = \inf_{ {\bm j}^{\perp} \in J^{(1)}(w) } \sup_{ {\bm J} \in \mathbb{R}^{|E|}} \left( - {\bm J}^T {\bm G} {\bm J}  + 2 {\bm J}^T {\bm j}^\perp \right) \, , \label{eqn:sig_vortex}
\end{align}
subjected to the constraint, 
${\bm \Phi}^{-1 \, T} {\bm J} \in \ker {\bm Q}$, 
imposed by the Lagrange multiplier vector ${\bm \nu}$. 
From Eq.~(\ref{eqn:imBTinKerD}), the replacement of ${\bm \Phi}^{-1 \, T} {\bm J}$ with 
${\bm B}^T {\bm A} \in \mathrm{im} {\bm B}^T$, 
where ${\bm A} \in \mathbb{R}^{|E(T^*)|}$ is the mesh current vector, leads to, 
\begin{align}
\frac{\langle \! \langle w \rangle \! \rangle^2}{\langle \! \langle w^2 \rangle \! \rangle}   &\leq \inf_{ {\bm j}^{\perp} \in J^{(1)}(w) } \sup_{ {\bm A} \in \mathbb{R}^{|E(T^*)|}} \left( - {\bm A}^T {\bm G}_2 {\bm A} + 2 {\bm A}^T {\bm B} {\bm \Phi} {\bm j}^\perp \right) \, , \label{eqn:exp_1chain}
\\
{\bm G}_2  &= {\bm B} {\bm \Phi} {\bm G} {\bm \Phi}^{T} {\bm B}^T \, . 
\end{align}
We find the maximizing ${\bm A}$ by solving 
${\bm G}_2 {\bm A} = {\bm B} {\bm \Phi} {\bm j}^\perp$. 
Then the right-hand side of Eq.~(\ref{eqn:exp_1chain}) becomes, 
\begin{align}
\inf_{ {\bm j}^{\perp} \in J^{(1)}(w) } \left( {\bm j}^{\perp \, T} {\bm \Phi}^T {\bm B}^T {\bm G}_{2}^{-1} {\bm B} {\bm \Phi} {\bm j}^\perp \right) \, . 
\end{align}
The replacement of ${\bm j}^{\perp} \in \ker {\bm D}$ with ${\bm B^T} {\bm f} \in \mathrm{im} {\bm B}^T$ leads to, 
\begin{align}
\frac{\langle \! \langle w \rangle \! \rangle^2}{\langle \! \langle w^2 \rangle \! \rangle} \leq & \inf_{ {\bm f} \in F(w)} \left( {\bm f}^T {\bm B} {\bm \Phi}^T {\bm B}^T {\bm G}_{2}^{-1} {\bm B} {\bm \Phi} {\bm B}^T {\bm f} \right) \, , \label{eqn:sig_2chain_}
\\
F(w) =& \left \{ {\bm f} \in \mathbb{R}^{|E(T^*)|} \middle | \langle \! \langle w \rangle \! \rangle = {\bm d}^T {\bm B}^T {\bm f} \right \} \, . \end{align}
Eventually, we reduce to a quadratic optimization problem in mesh currents $f_c$ ($c \in E(T^*)$) subjected to a linear constraint. 
In contrast to the mixed and pseudo-entropy bounds obtained by fixing ${\bm n}={\bm n}^{\rm st}$, this expression accounts for fluctuations of sojourn times at all nodes.

The inequalities (\ref{eqn:exp_1chain}) and (\ref{eqn:sig_2chain_}) result from the potential reduction of the parameter space as indicated by the inclusion relation (\ref{eqn:imBTinKerD}). 
However, one can deduce the equalities hold in the inequalities: 
The dimension of the image of ${\bm B}^T$, the cyclomatic number~\cite{Bryant1961}, is  
$\mathrm{dim} \, \mathrm{im} {\bm B}^T = \mathrm{rank} {\bm B} = |E(T^*)|$. 
By utilizing the equality $\mathrm{rank} {\bm D} = |V|-1 =|E(T)|$ for a connected graph,
$\mathrm{dim} \, \mathrm{ker} {\bm D} = |E| - \mathrm{rank} {\bm D} = |E| -|E(T)|$. 
Since 
$|E(T)|+|E(T^*)|=|E|$~\cite{Bryant1961}, 
we obtain 
$\mathrm{dim} \, \mathrm{im} {\bm B}^T = \mathrm{dim} \, \mathrm{ker} {\bm D}$, 
which, combined with the inclusion relation (\ref{eqn:imBTinKerD}), implies 
$\mathrm{im} {\bm B}^T = \mathrm{ker} {\bm D}$. 
Then the equality holds in Ineq.~(\ref{eqn:sig_2chain_}). 
Since 
$\mathrm{rank} {\bm D} = \mathrm{rank} {\bm Q}$, 
one can also deduce $\mathrm{im} {\bm B}^T = \mathrm{ker} {\bm Q}$, 
which implies that the equality holds in Ineq.~(\ref{eqn:exp_1chain}). 
\revis{
In this way, we obtain our main result, 
\begin{align}
\frac{\langle \! \langle w \rangle \! \rangle^2}{\langle \! \langle w^2 \rangle \! \rangle} = & \inf_{ {\bm f} \in F(w)} \left( {\bm f}^T {\bm B} {\bm \Phi}^T {\bm B}^T {\bm G}_{2}^{-1} {\bm B} {\bm \Phi} {\bm B}^T {\bm f} \right) \, . \label{eqn:sig_2chain}
\end{align}
A deviation from the optimal point results in a smaller second cumulant (see Supplemental Material~\cite{K5}), which serves as a lower bound.}

\modif{
Comments on the variational principles (\ref{eqn:Sigma_w}) and (\ref{eqn:sig_2chain}) are in order. 
Equation (\ref{eqn:Sigma_w}) is general but requires minimizing $|E|+|V|$ parameters under three constraints given by Eqs. (\ref{eqn:set_jw1}) and (\ref{eqn:set_p1}). 
Although (\ref{eqn:sig_2chain}) is limited to systems with uniform transition rates (\ref{eqn:uniform}), it only requires minimizing $|E(T^*)|$ parameters under one constraint. 
Therefore, (\ref{eqn:sig_2chain}) is more suitable for systems with few cycles, as will be demonstrated in the next section~\ref{sec:examples}.
}

\section{application: Brownian computation}
\label{sec:examples}

\begin{figure}[ht]
\begin{center}
\includegraphics[width=0.9 \columnwidth]{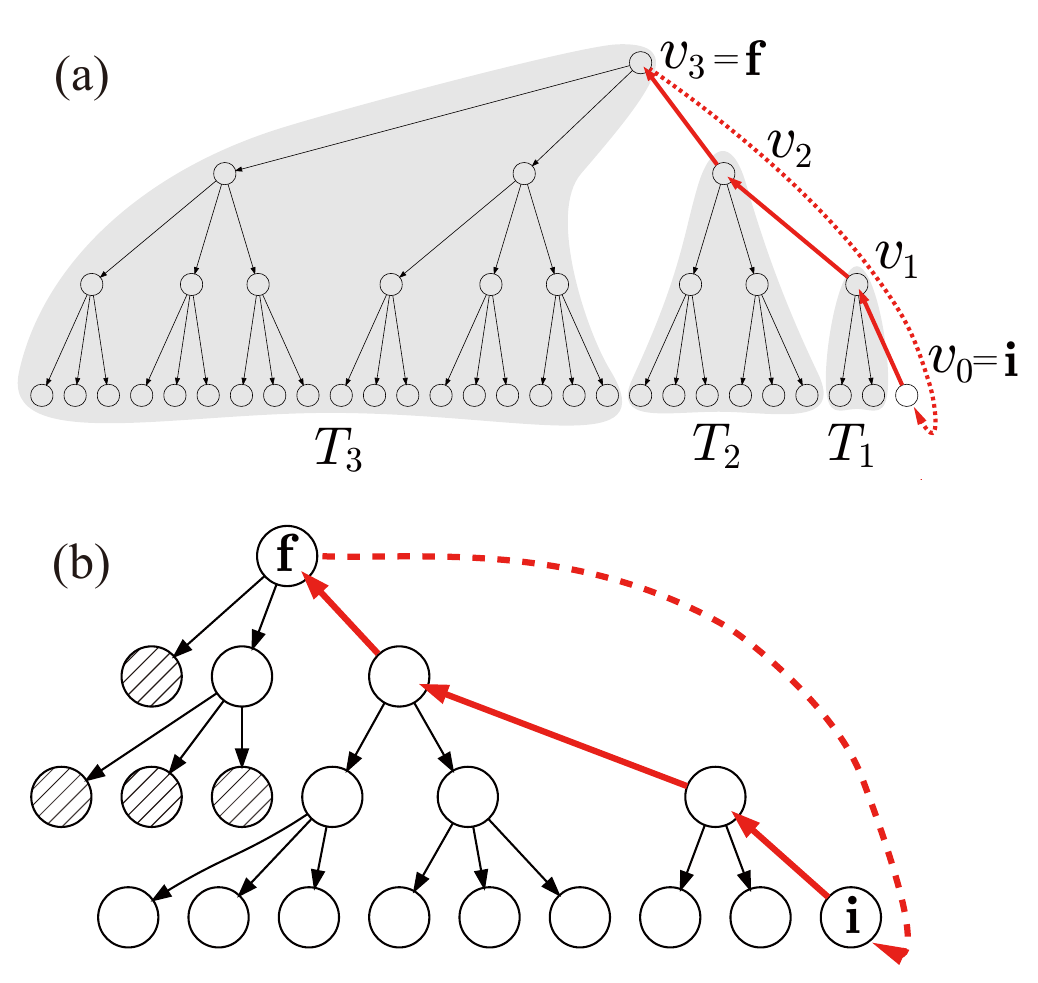}
\caption{
Graphs for schematic Brownian computation plus reset process. 
Solid arcs constitute the directed rooted spanning tree rooted at $v_0={\rm i}$, which is the start state. 
In each panel, thick solid arcs indicate the intended computation path from the start state $v_0={\rm i}$ to the output state $v_\ell={\rm f}$. 
The dotted arc is the chord indicating the unidirectional reset process. 
In panel (a), shaded subtrees represent extraneous predecessors of the logical states at the root $v_d$ ($d=1, \cdots, \ell$). 
}
\label{fig:GTree}
\end{center}
\end{figure}

We apply (\ref{eqn:sig_2chain}) to the Brownian computation process schematically depicted as tree graphs (see Fig.10 of Ref.~\cite{Bennett1982}). 
Figures~\ref{fig:GTree} (a) and (b) show such graphs whose nodes represent logical states. 
In each panel, the bottom nodes are possible input states, and the top node corresponds to the output state. 
The computation proceeds from bottom to top, and the branching represents the logical irreversibility. 
In each panel, solid arcs constitute a directed rooted spanning tree $T=(V,E(T))$. 
The computation starts from the start state, $v_0={\rm i}$, which we take as the root, to the final state $v_\ell={\rm f}$. 
There is an intended computation path with length $\ell$, $v_0 \to v_1 \to \cdots \to v_\ell$ [thick solid arcs in panels (a) and (b)], 
\begin{align}
P_{v_\ell \leftarrow v_0} = \left( t_{\ell}, t_{\ell-1}\cdots, t_1 \right) \, , 
\end{align}
where $t_d=(v_{d} \leftarrow v_{d-1})$. 
From each node on the intended computation path, a subtree $T_d$ rooted at the node $v_d$ grows [shaded parts in panel (a)]. 
The set of nodes is then $V = \bigcup_{d=0}^\ell V(T_{d})$, where $V(T_0)=\{ v_0 \}$. 
The nodes, except for the roots of the subtrees, represent extraneous predecessors; leaf nodes are either possible input states [bottom nodes in panels (a) and (b)] or `garden-of-Eden' states with no predecessors [hatched nodes in panel (b)]~\cite{Bennett1982}. 
All solid arcs are for bidirectional processes:  
\begin{align}
E(T) = E_{\rm b} = \bigcup_{d=0}^\ell E(T_{d}) \cup \{ t_1, \cdots, t_{\ell}\}  \, ,
\end{align}
where $E(T_{0}) = \emptyset$. 

The dotted arc in each panel is a chord $c$ representing the unidirectional reset process. 
The intended computation path and the reset path constitute a cycle: 
\begin{align}
C_{c}=(t_{\ell}, \cdots, t_1, c) \, .
\end{align}
The reset current is measured at the chord $d_e = \delta_{e,c}$. 
We assume that there is no other chord, i.e., $E(T^*) = E_{\rm uni} = \{ c \}$. 
Consequently, no free parameter exists in (\ref{eqn:sig_2chain}). 

We focus on the signal-to-noise ratio (SNR) of the probability distribution of the computation time, which we define as the first passage time~\cite{utsumi2022computation, Utsumi2023}, being upper bounded by TUR~\cite{GingrichPRL2017, Garrahan2017, PalPRR2021,utsumi2022computation}. 
The Fano factor of reset current and the SNR for $W$ resets are related as, 
\modif{$S/N \approx \sqrt{ W \langle \! \langle w \rangle \! \rangle / \langle \! \langle w^2 \rangle \! \rangle}$}
~\cite{utsumi2022computation}. 

From (\ref{eqn:sig_2chain}), the lower bound of the Fano factor is obtained as (Appendix~\ref{app:det_cal_exact_Fano_LB}), 
\begin{align}
\frac{ \langle \! \langle w^2 \rangle \! \rangle }{ \langle \! \langle w \rangle \! \rangle } =& \sum_{d=0}^{\ell} (2 \ell - 2 d +1) \left( \frac{{\mathcal N}_d}{\sum_{d=0}^\ell {\mathcal N}_d} \right)^2 
\nonumber \\ &+ \sum_{d=0}^\ell 2  (\ell-d+1) \sum_{e \in E(T_d)} \left( \frac{|V(T_{\partial^- e})|}{\sum_{d=0}^\ell {\mathcal N}_d} \right)^2 \, , 
\label{eqn:exact_Fano_LB}
\end{align}
where ${\mathcal N}_d = \sum_{d'=0}^d |V(T_{d'})|$ is monotonically increasing in $d$. 
The bound depends on the detailed structure of subtrees $T(v_d)$. 
In the following, we present analytic expressions for two examples. 

The first example is the model of Brownian logically reversible Turing machine (RTM)~\cite{utsumi2022computation, Utsumi2023}, in which extraneous predecessors are absent, 
$|V(T_{d})|=1$ and $|V(T_{\partial^- e})|=0$ ($e \in E(T_{d})$) for $d=0,\cdots, \ell=\tilde{\ell}-1$.
Then, ${\mathcal N}_d = d+1$ and thus the right-hand side of (\ref{eqn:exact_Fano_LB}) becomes, 
\begin{align}
\frac{2 (1+\tilde{\ell}+\tilde{\ell}^2)}{3 \tilde{\ell}(1+\tilde{\ell})} \, .
\label{eqn:Fano_1aryTree}
\end{align}
For $\ell \gg 1$, it approaches, $2/3$ as was obtained before~\cite{utsumi2022computation}. 

The second example is the case where the directed rooted spanning tree forms a complete $\alpha$-ary tree ($\alpha \geq 2$) if we reverse arcs of the intended computation path [Fig.~\ref{fig:GTree} (a)] 
\modif{(The complete $\alpha$-ary tree is a $\alpha$-ary tree in which all leaves have the same depth and all internal nodes have degree $\alpha$~\cite{Cormen_Thomas_H_2022-04-05})}. 
This case, the right-hand side of (\ref{eqn:exact_Fano_LB}) becomes (Appendix~\ref{app:Fano_a_aryTree}), 
\begin{align}
\frac{\alpha (\alpha^{\tilde{\ell}}-1)(4+\alpha+\alpha^{\tilde{\ell}+1}) + \tilde{\ell} (\alpha-1) (1+\alpha+4 \alpha^{\tilde{\ell}+1}) }{(\tilde{\ell} (1 - \alpha) + \alpha (\alpha^{\tilde{\ell}} -1 ))^2} \, .
\label{eqn:Fano_a_aryTree}
\end{align}
The lower bound approaches $1$ for $\ell \gg 1$. 
This behavior is reminiscent of the SNR observed for the token-based Brownian circuit~\cite{utsumi2022computation}. 
The weaker bound, which follows from (\ref{eqn:exact_Fano_LB}),  
\begin{align}
\frac{ \langle \! \langle w^2 \rangle \! \rangle }{ \langle \! \langle w \rangle \! \rangle } \geq  \sum_{d=0}^{\ell}  \left( \frac{{\mathcal N}_d}{\sum_{d'=0}^\ell {\mathcal N}_{d'}} \right)^2 \, , \label{eqn:qualitative_LB}
\end{align}
explains this behavior: 
The right-hand side approaches 1 when the extraneous predecessors are concentrating on the last subtree $T_{\ell}$, $|V(T_{\ell})| \approx |V|$. 
This condition is the case for the complete $\alpha$-ary tree-like graph, 
$|V(T_{\ell})|/|V| \approx 1-1/\alpha$ for $\ell \gg 1$. 

\begin{figure}[ht]
\begin{center}
\begin{minipage}[b]{0.9 \columnwidth}
\centering      
\includegraphics[width=1 \columnwidth]{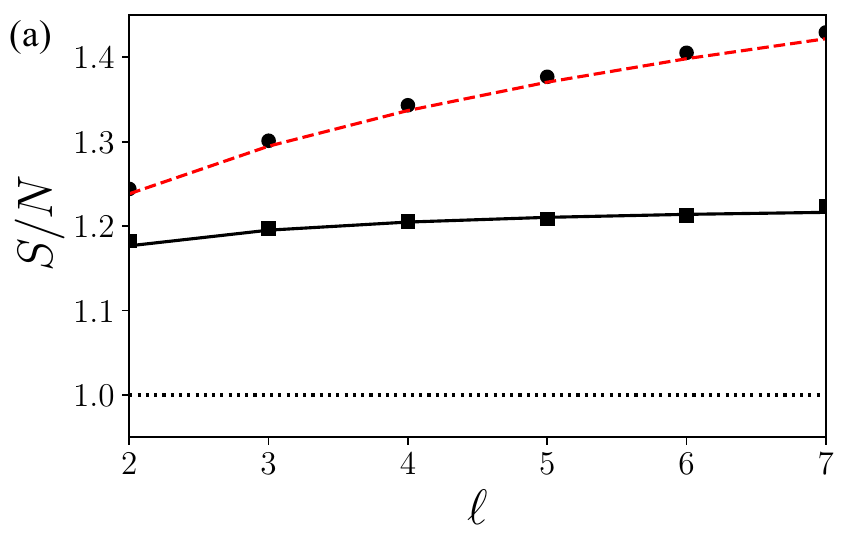} \\
\includegraphics[width=1 \columnwidth]{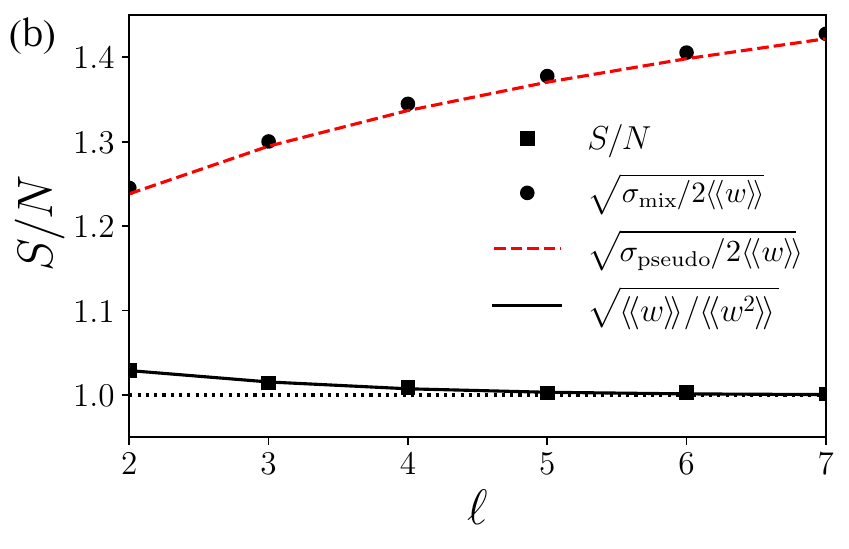} \\
\includegraphics[width=1 \columnwidth]{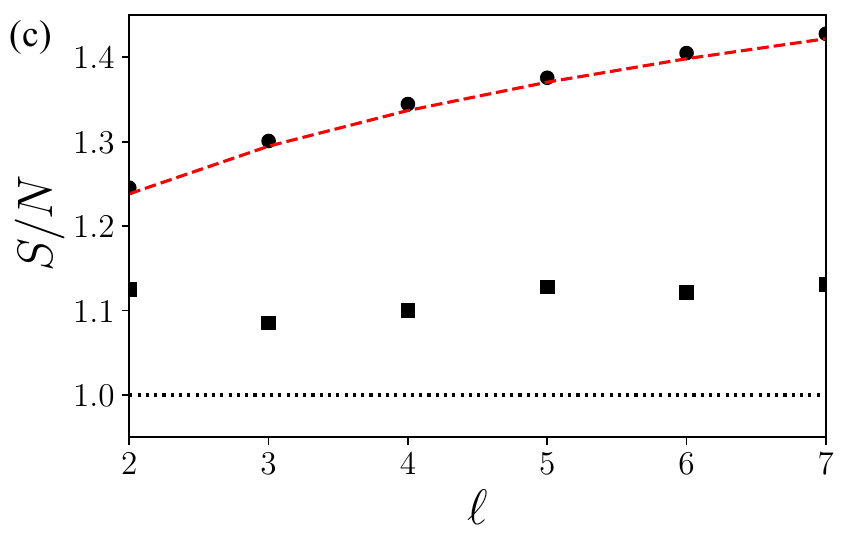}
\end{minipage}
\caption{
Plots of the signal-to-noise ratio $S/N$ versus the length of the intended computation path for 
\modif{(a)} the Brownian RTM, \modif{(b)} the complete 3-ary tree-like graph, Fig.~\ref{fig:GTree} (a), and \modif{(c)} the full 3-ary tree-like graph, Fig.~\ref{fig:GTree} (b) 
\modif{(The full $\alpha$-ary tree is a $\alpha$-ary tree in which each node is either a leaf or has degree $\alpha$~\cite{Cormen_Thomas_H_2022-04-05})}. 
The results are for a single reset ($W=1$). 
Filled squares are numerical results obtained by the Gillespie algorithm with $10^5$ samples. 
The numerical results \modif{in panels (a) and (b) are well fitted by the analytic expressions, Eqs.~(\ref{eqn:Fano_1aryTree}) and (\ref{eqn:Fano_a_aryTree})} (solid lines). 
The mixed-bound~(\ref{eqn:ana_mix}) (filled circles) and the pseudo-entropy bound~(\ref{eqn:ana_pseudo}) (dashed line) are shown in all panels. 
}
\label{fig:Tree}
\end{center}
\end{figure}

Figure \ref{fig:Tree} shows the SNR for a single reset ($W=1$) versus the length of the intended computation path $\ell$. 
\modif{Figures \ref{fig:Tree}(a) and (b) are for the Brownian RTM and for the complete 3-ary tree-like graph shown in Fig.~\ref{fig:GTree} (a), respectively.}
In both panels, the analytic \modif{expressions}, Eqs.~(\ref{eqn:Fano_1aryTree}) and (\ref{eqn:Fano_a_aryTree}) (solid lines) fit the numerical results (filled squares) obtained by the Gillespie algorithm. 
The SNR approaches $\sqrt{3/2}$ in the top panel, while in the middle panel, the SNR quickly approaches 1, i.e., the logical irreversibility degrades the SNR.

In each panel, filled circles and the dashed line indicate the mixed and pseudo-entropy bounds, (\ref{eqn:mix}) and (\ref{eqn:pseudo}),  
\begin{align}
\modif{ \frac{  \sigma_{\rm mix} }{ 2 \langle \! \langle w \rangle \! \rangle  } } =& \sum_{d=0}^{\ell-1} \frac{1}{2} \ln \frac{n_{d}^{\rm st}}{n_{d+1}^{\rm st}} +1 = \frac{1}{2} \ln \tilde{\ell} +1 \, , \label{eqn:ana_mix}
\\
\modif{ \frac{ \sigma_{\rm pseudo} }{ 2 \langle \! \langle w \rangle \! \rangle } } =& \sum_{d=0}^{\ell} \frac{\langle \! \langle w \rangle \! \rangle}{2 \ell -2 d +1} = \ln 2 + \frac{1} {2} H_{\ell+1/2} \nonumber \\ \approx& \ln 2 + \frac{\gamma} {2} + \frac{\ln \tilde{\ell}}{2} \, . \label{eqn:ana_pseudo}
\end{align}
Here $H_n$ is the $n$-th harmonic number, and $\gamma$ is the Euler's constant. 
Since the steady-state current flows along the cycle, the entropy production depends only on the length of the intended computation path $\ell$. 
It is independent of the number of the extraneous predecessors, which is large, e.g., there are $|V|-|C_c|=3272$ extraneous states for the complete 3-ary tree-like graph with $\ell=7$ corresponding to \modif{Fig.~\ref{fig:Tree} (b)}. 
\modif{In general, for a complete $\alpha$-ary tree-like graph, although there are $\sim \alpha^\ell$ nodes, the two bounds diverge as $\sim \ln \ell$ independently of $\alpha$. In other words, the two bounds fail to capture the logical irreversibility and become logarithmically weaker for a longer intended computation path.}

\modif{Figure~\ref{fig:Tree}(c)} corresponds to the 3-ary tree-like graph, Fig.~\ref{fig:GTree} (b). 
The SNR is bigger than one and smaller than mixed and pseudo-entropy bounds. 
Note that the average computation time itself, which is the reciprocal of the average reset current 
$1/\langle \! \langle w \rangle \! \rangle = \sum_{d=0}^\ell (\ell-d+1) |V(T_d)|/\Gamma$, increases with the number of extraneous states. 
We also note that the schematic Brownian computation models discussed here discard certain features of potential Brownian computation models: 
The token-based Brownian computation model~\cite{utsumi2022computation} is concurrent and thus should contain many cycles in its graph.
These cycles change the steady-state probabilities and alter the path-length dependence of mixed and pseudo-entropy bounds.

\section{Conclusion}
\label{sec:conclusion}

Starting from the level 2.5 large deviation function, we derive a lower bound of the Fano factor expressed as a quadratic optimization problem in mesh currents subjected to a linear constraint. 
By the duality transformation and exploiting the root-to-node path matrix, we effectively integrate out fluctuations of sojourn times at all nodes when rates for bidirectional and unidirectional transitions are uniform.
\modif{
The main result, (\ref{eqn:sig_2chain}), highlights a connection between current fluctuations and the network topology, as well as the time-symmetric activity out of equilibrium.}
The bound applied to the schematic Brownian computation plus reset process shows that the logical irreversibility reduces the signal-to-noise ratio of the fluctuating computation time. 
It is contrasted with the mixed and pseudo-entropy bounds, which are independent of the logical irreversibility and become weaker logarithmically in the length of the intended computation path. 

\modif{
Although the present paper is limited to systems with uniform transition rates, they include, e.g., certain types of Brownian computation or search without heat production, and are thus of physical relevance. 
The generalization to non-uniform transition rates will be discussed in the future.
}

\begin{acknowledgments}
We thank Satoshi Nakajima for proving the inequality (\ref{eqn:nakajima_ineq}). 
This work was supported by JSPS KAKENHI Grants No. 18KK0385, No. 20H01827, No. 20H05666, and No. 24K00547, and JST, CREST Grant Number JPMJCR20C1, Japan.
\end{acknowledgments}

\appendix

\section{Derivations of Eq.~(\ref{eqn:rate_func_1})}
\label{app:derivation_rate_function}

\modif{
At the maximum point satifying Eq.~(\ref{eqn:saddle_point}), 
the eigenvalue $\Lambda = \sum_{v,v'\in V} u^L_{v} L_{v,v'} u^R_{v'}$ reads,
\begin{align}
\Lambda =&
\sum_{e \in E_{\rm m}} \Gamma_e \left( e^{x_e } u^L_{\partial^- e} u^R_{\partial^+ e} 
- u^L_{\partial^+ e} u^R_{\partial^+ e} \right) + \sum_{v \in V} y_v u^L_{v} u^R_{v} \\
=&
\sum_{e \in A} \left( w_e - \Gamma_e n_{\partial^+ e} \right) + \sum_{v \in V} y_v n_v 
\, . \label{eqn:eigenvalue_at_sp}
\end{align}
Here, to obtain Eq.~(\ref{eqn:eigenvalue_at_sp}), we used $e^{x_e} = {w_e}/(u^L_{\partial^- e} u^R_{\partial^+ e})$, which is equivalent to Eq.~(\ref{eqn:saddle_point}). 
The factor $\sum_{e \in E_{\rm m}} x_e w_e$ at the maximum point reads, 
\begin{align}
\sum_{e \in E_{\rm m}} x_e w_e = 
\sum_{e \in E_{\rm m}} w_e \ln \frac{w_e}{a_e} - \sum_{v \in V} \sum_{e \in E_{\rm m}} w_e D_{v,e} \ln u^L_{v} 
\label{eqn:x_w_at_sp}
\, ,
\end{align}
where the second term of the right-hand side vanishes because of the KCL (\ref{eqn:KCL_g}). 
Then by combining Eqs.~(\ref{eqn:eigenvalue_at_sp}), (\ref{eqn:x_w_at_sp}) and (\ref{eqn:Omega}), we obtain Eq.~(\ref{eqn:rate_func_1}). 
}

\section{Proof of the inequality (\ref{eqn:nakajima_ineq})}
\label{app:nakajima_ineq}

Let us introduce a function, 
\begin{align}
f(x) = \psi(x)  + \frac{(x-1)^2}{2} - \frac{(x-1)^3}{6} + \frac{(x-1)^4}{3} \, . 
\end{align}
The first and second derivatives are, 
\begin{align}
f'(x) =& -\ln x + \frac{(x-1)(8 x^2 -19 x+17)}{6} \, , \\
f''(x) =& \frac{(x-1)^2 (4x-1)}{x} \, . 
\end{align}
Since, $f''(x)<0$ for $0<x<1/4$ and $f''(x) \geq 0$ for $1/4 \leq x$, 
$f'(x)$ is monotonically decreasing for $0<x<1/4$ and weakly increasing for $1/4 \leq x$. 
The zeros of $f'(x)$ are $x=1$ and $x=x^*$; 
Since $\lim_{x \to +0} f'(x) = \infty$ and $f'(1/4)=\ln 4-51/32 = -0.207 \cdots <0$, there exists $x^*$ such that $0<x^*<1/4$ and $f'(x^*)=0$. 
Then, $f'(x) \leq 0$ for $x^* \leq x \leq 1$, and $f'(x)>0$ for $0<x<x^*$ and $1<x$. 
Therefore, a local maximum and a local minimum exist at $x=x^*$ and $x=1$, respectively.  
Since $\lim_{x \to +0} f(x) = +0$ and the local minimum is $f(1)=0$, we conclude 
$f(x) \geq 0$ for $x>0$, which proves (\ref{eqn:nakajima_ineq}).

\section{Proofs of Eqs.~(\ref{eqn:cutsetmatrix}) and (\ref{eqn:DBT0})}

\label{app:cutsetmatrix}

Equation~(\ref{eqn:cutsetmatrix}) is calculated as, 
$\left( {\bm Q} \right)_{t,e} = \mathbbm{1}_{P_{\partial^- e \leftarrow v_0}}(t) - \mathbbm{1}_{P_{\partial^+ e \leftarrow v_0}}(t)$. 
If $e \in E(T)$, it is 1 only when $t=e$ and 0 otherwise. 
If $e \in E(T^*)$, it is $-1$($1$), when $t$($-t$) is in the cycle ${C}_e$. 
Therefore, 
\begin{align}
\left( {\bm Q} \right)_{t,e} &= \left \{ \begin{array}{cc} 
\delta_{t,e} & ( e \in E(T) ) \\
\mathbbm{1}_{C_e}(-t) - \mathbbm{1}_{C_e}(t) =F_{t,e}  & ( e \in E(T^*) ) 
\end{array} \right. \, , 
\end{align} 
which proves Eq.~(\ref{eqn:cutsetmatrix})~\cite{Takahashi1969}. 

We introduce $\delta^\pm$, the dual of $\partial^\pm$, that maps $V$ to $E$ as, 
$\delta^\pm v = \left \{ e \middle | \partial^\pm e =v \wedge e \in E \right \}$. 
By using this, the incidence matrix is expressed as 
$ \left( {\bm D}^\pm \right)_{v,e} = \mathbbm{1}_{\delta^\pm v} (e)$. 
Then $\left( {\bm D}^\pm {\bm B}^T \right)_{v,c}$ is calculated as, 
\begin{align}
\left( {\bm D}^\pm {\bm B}^T \right)_{v,c} = \sum_{e \in C_c} \mathbbm{1}_{\delta^\pm v}(e)\, , \label{eqn:DB1}
\end{align} 
if the cycle $C_c$ goes along the directions of twigs of $T$, and 
\begin{align}
\left( {\bm D}^\pm {\bm B}^T \right)_{v,c} 
= \mathbbm{1}_{\delta^\pm v} (c) - \sum_{t' \in C_c - \{ c \}} \mathbbm{1}_{\delta^\pm v}(-t') \, , \label{eqn:DB2}
\end{align} 
if $C_c$ goes in the direction opposite to $T$. 
Here we write a path obtained from $C_c$ by removing $c$ as $C_c - \{ c \}$. 
For both equations, (\ref{eqn:DB1}) and (\ref{eqn:DB2}), we obtain $\left( {\bm D} {\bm B}^T \right)_{v,c}=0$.
\modif{
In general, 
\begin{align}
\left( {\bm D}^\pm {\bm B}^T \right)_{v,c} 
=& \mathbbm{1}_{\delta^\pm v} (c) 
+ 
\sum_{t \in (C_c-\{c\}) \cap E(T)} 
\mathbbm{1}_{\delta^\pm v} (t)
\nonumber \\
&- 
\sum_{t' \in (C_c-\{c\}) \cap \overline{E(T)} } 
\mathbbm{1}_{\delta^\pm v} (-t')
\, , \label{eqn:DB3}
\end{align} 
where $\overline{E(T)}$ is the set of reversed twigs, 
also leads $\left( {\bm D} {\bm B}^T \right)_{v,c}=0$ and proves Eq.~(\ref{eqn:DBT0}). 
}

\section{${\bm \Phi}$ matrix}
\label{app:Phi}

We write a unit vector as, 
${\bm e}_e^T = \begin{bmatrix} {\bm e}_e(T)^T & {\bm e}_{e}(T^*)^T \end{bmatrix}$, 
where 
${\bm e}_e(T) \in \mathbb{R}^{|E(T)|}$
and
${\bm e}_e(T^*) \in \mathbb{R}^{|E(T^*)|}$. 
Accordingly, 
\begin{align}
\begin{bmatrix} {\bm \Omega}(T) \\ {\bm \Omega}(T^*) \end{bmatrix} 
= \begin{bmatrix} 
\sum_{e' \in E_{\rm uni} \cap E(T)} {\bm e}_{e'}(T) {\bm \Omega}_{e'}^T \\
\sum_{e' \in E_{\rm uni} \cap E(T^*)} {\bm e}_{e'}(T^*) {\bm \Omega}_{e'}^T 
\end{bmatrix} \, .
\end{align} 
By substituting them into Eq.~(\ref{eqn:Phi_matrix}), we obtain, 
\begin{align}
{\bm \Phi} = \begin{bmatrix} \left( {\bm I}_{|E(T)|} + {\bm \Omega}(T) \right)^{-1} & {\bm 0} \\ 
-{\bm \Omega}(T^*) \left( {\bm I}_{|E(T)|} + {\bm \Omega}(T) \right)^{-1}  & {\bm I}_{|E(T^*)|} \end{bmatrix} \, .
\end{align} 

\section{Derivations of Eq.~(\ref{eqn:exact_Fano_LB})}
\label{app:det_cal_exact_Fano_LB}

By exploiting Eq.~(\ref{eqn:KH}), the node state probability in the steady-state is, 
\begin{align}
n_v^{\rm st} = (\ell-d+1)/Z \, , \;\;\;\; (v \in V(T_{d})) \, .
\end{align}

The $1 \times |E|$ cycle matrix is $B_{c,e}=\mathbbm{1}_{C_c}(e)$. 
The nonzero components of $|E| \times |E|$ matrix ${\bm \Phi}$ is
${\Phi}_{e,e} = 1$ for $e \in E$ and 
${\Phi}_{c,t} = -|V(T_{\partial^- t})|/|V| $
for $t \in E(T)$ and $c \in E(T^*)$ (Appendix \ref{app:Phi}). 
The mesh current satisfying the constraint is 
${\bm f}= \langle \! \langle w \rangle \! \rangle = \Gamma g_c$. 

The number of node of subtree rooted at $\partial^- t_d$ is $|V(T_{\partial^- t_d })| = \sum_{d'=d}^\ell |V(T_{d'})|$. 
By using this, we obtain, 
\begin{align}
\sum_{e \in C_c} {\Phi}_{e,t_d} = 1 - |V(T_{\partial^- t_d })|/|V| = {\mathcal N}_{d-1}/|V| \, .
\end{align}
The relations mentioned above lead to, 
\begin{align}
{\bm B} {\bm \Phi} {\bm B}^T =& \sum_{d=1}^{\ell} \sum_{e \in C_c}  {\Phi}_{e,t_d} + 1 = \sum_{d=0}^\ell \frac{ {\mathcal N}_d }{|V|} \, , \label{eqn:numerator}
\\
\frac{ {\bm G}_2 }{\langle \! \langle w \rangle \! \rangle}
=&
\sum_{d=1}^{\ell} \frac{g_{t_d}}{g_c} \left( \sum_{e' \in C_c}  \Phi_{e',t_d} \right)^2
+ \left( \sum_{e' \in C_c}  \Phi_{e',c} \right)^2 \nonumber \\ &
+
\sum_{d=0}^\ell
\sum_{e \in E(T_d)} \frac{g_{e}}{g_c} \left( \sum_{e' \in C_c}  \Phi_{e',e} \right)^2
\\
=&
\sum_{d=0}^{\ell} (2 \ell - 2 d +1) \left( \frac{ {\mathcal N}_d }{|V|} \right)^2 + \sum_{d=0}^\ell 2  (\ell-d+1) \nonumber \\ & \times \sum_{e \in E(T_d)} \left( \frac{|V(T_{\partial^- e})|}{|V|} \right)^2 \, ,
\label{eqn:denom}
\end{align}
where we used 
$g_e/g_c=2 n^{\rm st}_{v_d}/n^{\rm st}_{v_\ell}$
and
$\sum_{e' \in C_c}  \Phi_{e',e} = \Phi_{c,e}$
for $e \in E(T_d)$. 
By combining Eqs.~(\ref{eqn:numerator}) and (\ref{eqn:denom}), 
the right-hand side of Eq.~(\ref{eqn:exact_Fano_LB}), 
${\bm G}_2 \langle \! \langle w \rangle \! \rangle/\left({\bm B} {\bm \Phi} {\bm B}^T {\bm f} \right)^2$ is obtained. 

\section{Derivations of Eq.~(\ref{eqn:Fano_a_aryTree})}
\label{app:Fano_a_aryTree}

Let $T(\alpha,h)$ a complete $\alpha$-ary tree of \modif{height} $h$. 
It contains 
$|V(T(\alpha,h))|=\sum_{h'=0}^h \alpha^{h'}=(\alpha^{h+1}-1)/(\alpha-1)$ nodes. 
In $T_d$, there are $N(T(\alpha,h))=\alpha^{d-h-1}(\alpha-1)$ ($h=0,\cdots,d-1$) such subtrees. 
Therefore, 
\begin{align}
|V(T_{d})| = 1+ \sum_{h=0}^{d-1} N(T(\alpha,h)) = \alpha^{d} \, .
\end{align}
By exploiting these relations, we obtain, 
\begin{align}
{\mathcal N}_d &= \frac{\alpha^{d+1} -1}{\alpha-1} \, \\
\sum_{d=0}^\ell {\mathcal N}_{d} &= \frac{1+\alpha^{\ell+2}+\ell-\alpha (\ell+2)}{(\alpha-1)^2} \, , 
\\ \sum_{e \in E(T_{d})} |V(T_{\partial^- e})|^2 =& \sum_{h=0}^{d-1} N(T(\alpha,h)) |V(T(\alpha,h))|^2 \nonumber \\
=& \frac{\alpha^{d} [ (1+2 d) (1-a)+a^{d+1}]-1 }{(\alpha-1)^2 } \, . 
\end{align}
By substituting them into the right-hand side of (\ref{eqn:exact_Fano_LB}), 
we obtain (\ref{eqn:Fano_a_aryTree}).

\bibliography{level25_resubmit}

\begin{thebibliography}{37}%
\makeatletter
\providecommand \@ifxundefined [1]{%
 \@ifx{#1\undefined}
}%
\providecommand \@ifnum [1]{%
 \ifnum #1\expandafter \@firstoftwo
 \else \expandafter \@secondoftwo
 \fi
}%
\providecommand \@ifx [1]{%
 \ifx #1\expandafter \@firstoftwo
 \else \expandafter \@secondoftwo
 \fi
}%
\providecommand \natexlab [1]{#1}%
\providecommand \enquote  [1]{``#1''}%
\providecommand \bibnamefont  [1]{#1}%
\providecommand \bibfnamefont [1]{#1}%
\providecommand \citenamefont [1]{#1}%
\providecommand \href@noop [0]{\@secondoftwo}%
\providecommand \href [0]{\begingroup \@sanitize@url \@href}%
\providecommand \@href[1]{\@@startlink{#1}\@@href}%
\providecommand \@@href[1]{\endgroup#1\@@endlink}%
\providecommand \@sanitize@url [0]{\catcode `\\12\catcode `\$12\catcode `\&12\catcode `\#12\catcode `\^12\catcode `\_12\catcode `\%12\relax}%
\providecommand \@@startlink[1]{}%
\providecommand \@@endlink[0]{}%
\providecommand \url  [0]{\begingroup\@sanitize@url \@url }%
\providecommand \@url [1]{\endgroup\@href {#1}{\urlprefix }}%
\providecommand \urlprefix  [0]{URL }%
\providecommand \Eprint [0]{\href }%
\providecommand \doibase [0]{https://doi.org/}%
\providecommand \selectlanguage [0]{\@gobble}%
\providecommand \bibinfo  [0]{\@secondoftwo}%
\providecommand \bibfield  [0]{\@secondoftwo}%
\providecommand \translation [1]{[#1]}%
\providecommand \BibitemOpen [0]{}%
\providecommand \bibitemStop [0]{}%
\providecommand \bibitemNoStop [0]{.\EOS\space}%
\providecommand \EOS [0]{\spacefactor3000\relax}%
\providecommand \BibitemShut  [1]{\csname bibitem#1\endcsname}%
\let\auto@bib@innerbib\@empty
\bibitem [{\citenamefont {Horowitz}\ and\ \citenamefont {Gingrich}(2020)}]{Horowitz2020}%
  \BibitemOpen
  \bibfield  {author} {\bibinfo {author} {\bibfnamefont {J.~M.}\ \bibnamefont {Horowitz}}\ and\ \bibinfo {author} {\bibfnamefont {T.~R.}\ \bibnamefont {Gingrich}},\ }\bibfield  {title} {\bibinfo {title} {Thermodynamic uncertainty relations constrain non-equilibrium fluctuations},\ }\href {https://doi.org/10.1038/s41567-019-0702-6} {\bibfield  {journal} {\bibinfo  {journal} {Nature Physics}\ }\textbf {\bibinfo {volume} {16}},\ \bibinfo {pages} {15} (\bibinfo {year} {2020})}\BibitemShut {NoStop}%
\bibitem [{\citenamefont {Shiraishi}(2023)}]{ShiraishiBook2023}%
  \BibitemOpen
  \bibfield  {author} {\bibinfo {author} {\bibfnamefont {N.}~\bibnamefont {Shiraishi}},\ }\href {https://doi.org/10.1007/978-981-19-8186-9} {\emph {\bibinfo {title} {An Introduction to Stochastic Thermodynamics: From Basic to Advanced}}}\ (\bibinfo  {publisher} {Springer Nature Singapore},\ \bibinfo {address} {Singapore},\ \bibinfo {year} {2023})\BibitemShut {NoStop}%
\bibitem [{\citenamefont {Wolpert}(2020)}]{Wolpert2020}%
  \BibitemOpen
  \bibfield  {author} {\bibinfo {author} {\bibfnamefont {D.~H.}\ \bibnamefont {Wolpert}},\ }\bibfield  {title} {\bibinfo {title} {Uncertainty relations and fluctuation theorems for bayes nets},\ }\href {https://doi.org/10.1103/PhysRevLett.125.200602} {\bibfield  {journal} {\bibinfo  {journal} {Phys. Rev. Lett.}\ }\textbf {\bibinfo {volume} {125}},\ \bibinfo {pages} {200602} (\bibinfo {year} {2020})}\BibitemShut {NoStop}%
\bibitem [{\citenamefont {Utsumi}\ \emph {et~al.}(2022)\citenamefont {Utsumi}, \citenamefont {Ito}, \citenamefont {Golubev},\ and\ \citenamefont {Peper}}]{utsumi2022computation}%
  \BibitemOpen
  \bibfield  {author} {\bibinfo {author} {\bibfnamefont {Y.}~\bibnamefont {Utsumi}}, \bibinfo {author} {\bibfnamefont {Y.}~\bibnamefont {Ito}}, \bibinfo {author} {\bibfnamefont {D.}~\bibnamefont {Golubev}},\ and\ \bibinfo {author} {\bibfnamefont {F.}~\bibnamefont {Peper}},\ }\bibfield  {title} {\bibinfo {title} {Computation time and thermodynamic uncertainty relation of brownian circuits},\ }\href@noop {} {\  (\bibinfo {year} {2022})},\ \Eprint {https://arxiv.org/abs/2205.10735} {arXiv:2205.10735 [cond-mat.stat-mech]} \BibitemShut {NoStop}%
\bibitem [{\citenamefont {Bennett}(1982)}]{Bennett1982}%
  \BibitemOpen
  \bibfield  {author} {\bibinfo {author} {\bibfnamefont {C.~H.}\ \bibnamefont {Bennett}},\ }\bibfield  {title} {\bibinfo {title} {The thermodynamics of computation—a review},\ }\href {https://doi.org/10.1007/BF02084158} {\bibfield  {journal} {\bibinfo  {journal} {International Journal of Theoretical Physics}\ }\textbf {\bibinfo {volume} {21}},\ \bibinfo {pages} {905} (\bibinfo {year} {1982})}\BibitemShut {NoStop}%
\bibitem [{\citenamefont {Wolpert}\ \emph {et~al.}(2023)\citenamefont {Wolpert}, \citenamefont {Korbel}, \citenamefont {Lynn}, \citenamefont {Tasnim}, \citenamefont {Grochow}, \citenamefont {Kardeş}, \citenamefont {Aimone}, \citenamefont {Balasubramanian}, \citenamefont {de~Giuli}, \citenamefont {Doty}, \citenamefont {Freitas}, \citenamefont {Marsili}, \citenamefont {Ouldridge}, \citenamefont {Richa}, \citenamefont {Riechers}, \citenamefont {Édgar Roldán}, \citenamefont {Rubenstein}, \citenamefont {Toroczkai},\ and\ \citenamefont {Paradiso}}]{wolpert2023stochastic}%
  \BibitemOpen
  \bibfield  {author} {\bibinfo {author} {\bibfnamefont {D.}~\bibnamefont {Wolpert}}, \bibinfo {author} {\bibfnamefont {J.}~\bibnamefont {Korbel}}, \bibinfo {author} {\bibfnamefont {C.}~\bibnamefont {Lynn}}, \bibinfo {author} {\bibfnamefont {F.}~\bibnamefont {Tasnim}}, \bibinfo {author} {\bibfnamefont {J.}~\bibnamefont {Grochow}}, \bibinfo {author} {\bibfnamefont {G.}~\bibnamefont {Kardeş}}, \bibinfo {author} {\bibfnamefont {J.}~\bibnamefont {Aimone}}, \bibinfo {author} {\bibfnamefont {V.}~\bibnamefont {Balasubramanian}}, \bibinfo {author} {\bibfnamefont {E.}~\bibnamefont {de~Giuli}}, \bibinfo {author} {\bibfnamefont {D.}~\bibnamefont {Doty}}, \bibinfo {author} {\bibfnamefont {N.}~\bibnamefont {Freitas}}, \bibinfo {author} {\bibfnamefont {M.}~\bibnamefont {Marsili}}, \bibinfo {author} {\bibfnamefont {T.~E.}\ \bibnamefont {Ouldridge}}, \bibinfo {author} {\bibfnamefont {A.}~\bibnamefont {Richa}}, \bibinfo {author} {\bibfnamefont {P.}~\bibnamefont {Riechers}}, \bibinfo {author} {\bibnamefont {Édgar Roldán}}, \bibinfo {author} {\bibfnamefont {B.}~\bibnamefont {Rubenstein}}, \bibinfo {author} {\bibfnamefont {Z.}~\bibnamefont {Toroczkai}},\ and\ \bibinfo {author} {\bibfnamefont {J.}~\bibnamefont {Paradiso}},\ }\bibfield  {title} {\bibinfo {title} {Is stochastic thermodynamics the key to understanding the energy costs of computation?},\ }\href@noop {} {\  (\bibinfo {year} {2023})},\ \Eprint {https://arxiv.org/abs/2311.17166} {arXiv:2311.17166 [cond-mat.stat-mech]} \BibitemShut {NoStop}%
\bibitem [{\citenamefont {Bryant}(1961)}]{Bryant1961}%
  \BibitemOpen
  \bibfield  {author} {\bibinfo {author} {\bibfnamefont {P.}~\bibnamefont {Bryant}},\ }\bibfield  {title} {\bibinfo {title} {The algebra and topology of electrical networks},\ }\href {https://digital-library.theiet.org/content/journals/10.1049/pi-c.1961.0030} {\bibfield  {journal} {\bibinfo  {journal} {Proceedings of the IEE - Part C: Monographs}\ }\textbf {\bibinfo {volume} {108}},\ \bibinfo {pages} {215} (\bibinfo {year} {1961})}\BibitemShut {NoStop}%
\bibitem [{\citenamefont {Branin}(1967)}]{Branin1967}%
  \BibitemOpen
  \bibfield  {author} {\bibinfo {author} {\bibfnamefont {F.}~\bibnamefont {Branin}},\ }\bibfield  {title} {\bibinfo {title} {Computer methods of network analysis},\ }\href {https://doi.org/10.1109/PROC.1967.6010} {\bibfield  {journal} {\bibinfo  {journal} {Proceedings of the IEEE}\ }\textbf {\bibinfo {volume} {55}},\ \bibinfo {pages} {1787} (\bibinfo {year} {1967})}\BibitemShut {NoStop}%
\bibitem [{\citenamefont {Takahashi}(1969)}]{Takahashi1969}%
  \BibitemOpen
  \bibfield  {author} {\bibinfo {author} {\bibfnamefont {H.}~\bibnamefont {Takahashi}},\ }\href@noop {} {\emph {\bibinfo {title} {Theory of Linear Lumped Parameter System I}}},\ \bibinfo {series} {Kisokougaku, Iwanami kouza}\ No.~\bibinfo {number} {6}\ (\bibinfo  {publisher} {Iwanami shoten},\ \bibinfo {address} {Tokyo},\ \bibinfo {year} {1969})\BibitemShut {NoStop}%
\bibitem [{\citenamefont {Rasmussen}\ \emph {et~al.}(2021)\citenamefont {Rasmussen}, \citenamefont {Christensen}, \citenamefont {Pedersen}, \citenamefont {Kristensen}, \citenamefont {B\ae{}kkegaard}, \citenamefont {Loft},\ and\ \citenamefont {Zinner}}]{RasmussenPRX2021}%
  \BibitemOpen
  \bibfield  {author} {\bibinfo {author} {\bibfnamefont {S.}~\bibnamefont {Rasmussen}}, \bibinfo {author} {\bibfnamefont {K.}~\bibnamefont {Christensen}}, \bibinfo {author} {\bibfnamefont {S.}~\bibnamefont {Pedersen}}, \bibinfo {author} {\bibfnamefont {L.}~\bibnamefont {Kristensen}}, \bibinfo {author} {\bibfnamefont {T.}~\bibnamefont {B\ae{}kkegaard}}, \bibinfo {author} {\bibfnamefont {N.}~\bibnamefont {Loft}},\ and\ \bibinfo {author} {\bibfnamefont {N.}~\bibnamefont {Zinner}},\ }\bibfield  {title} {\bibinfo {title} {Superconducting circuit companion---an introduction with worked examples},\ }\href {https://doi.org/10.1103/PRXQuantum.2.040204} {\bibfield  {journal} {\bibinfo  {journal} {PRX Quantum}\ }\textbf {\bibinfo {volume} {2}},\ \bibinfo {pages} {040204} (\bibinfo {year} {2021})}\BibitemShut {NoStop}%
\bibitem [{\citenamefont {Utsumi}\ \emph {et~al.}(2023)\citenamefont {Utsumi}, \citenamefont {Golubev},\ and\ \citenamefont {Peper}}]{Utsumi2023}%
  \BibitemOpen
  \bibfield  {author} {\bibinfo {author} {\bibfnamefont {Y.}~\bibnamefont {Utsumi}}, \bibinfo {author} {\bibfnamefont {D.}~\bibnamefont {Golubev}},\ and\ \bibinfo {author} {\bibfnamefont {F.}~\bibnamefont {Peper}},\ }\bibfield  {title} {\bibinfo {title} {Thermodynamic cost of brownian computers in the stochastic thermodynamics of resetting},\ }\href {https://doi.org/10.1140/epjs/s11734-023-00981-8} {\bibfield  {journal} {\bibinfo  {journal} {The European Physical Journal Special Topics}\ }\textbf {\bibinfo {volume} {232}},\ \bibinfo {pages} {3259} (\bibinfo {year} {2023})}\BibitemShut {NoStop}%
\bibitem [{\citenamefont {Bertini}\ \emph {et~al.}(2015{\natexlab{a}})\citenamefont {Bertini}, \citenamefont {Faggionato},\ and\ \citenamefont {Gabrielli}}]{BertiniAIHP2015}%
  \BibitemOpen
  \bibfield  {author} {\bibinfo {author} {\bibfnamefont {L.}~\bibnamefont {Bertini}}, \bibinfo {author} {\bibfnamefont {A.}~\bibnamefont {Faggionato}},\ and\ \bibinfo {author} {\bibfnamefont {D.}~\bibnamefont {Gabrielli}},\ }\bibfield  {title} {\bibinfo {title} {{Large deviations of the empirical flow for continuous time Markov chains}},\ }\href {https://doi.org/10.1214/14-AIHP601} {\bibfield  {journal} {\bibinfo  {journal} {Annales de l'Institut Henri Poincaré, Probabilités et Statistiques}\ }\textbf {\bibinfo {volume} {51}},\ \bibinfo {pages} {867 } (\bibinfo {year} {2015}{\natexlab{a}})}\BibitemShut {NoStop}%
\bibitem [{\citenamefont {Bertini}\ \emph {et~al.}(2015{\natexlab{b}})\citenamefont {Bertini}, \citenamefont {Faggionato},\ and\ \citenamefont {Gabrielli}}]{BERTINI2015}%
  \BibitemOpen
  \bibfield  {author} {\bibinfo {author} {\bibfnamefont {L.}~\bibnamefont {Bertini}}, \bibinfo {author} {\bibfnamefont {A.}~\bibnamefont {Faggionato}},\ and\ \bibinfo {author} {\bibfnamefont {D.}~\bibnamefont {Gabrielli}},\ }\bibfield  {title} {\bibinfo {title} {Flows, currents, and cycles for markov chains: Large deviation asymptotics},\ }\href {https://doi.org/https://doi.org/10.1016/j.spa.2015.02.001} {\bibfield  {journal} {\bibinfo  {journal} {Stochastic Processes and their Applications}\ }\textbf {\bibinfo {volume} {125}},\ \bibinfo {pages} {2786} (\bibinfo {year} {2015}{\natexlab{b}})}\BibitemShut {NoStop}%
\bibitem [{\citenamefont {Barato}\ and\ \citenamefont {Chetrite}(2015)}]{Barato2015}%
  \BibitemOpen
  \bibfield  {author} {\bibinfo {author} {\bibfnamefont {A.~C.}\ \bibnamefont {Barato}}\ and\ \bibinfo {author} {\bibfnamefont {R.}~\bibnamefont {Chetrite}},\ }\bibfield  {title} {\bibinfo {title} {A formal view on level 2.5 large deviations and fluctuation relations},\ }\href {https://doi.org/10.1007/s10955-015-1283-0} {\bibfield  {journal} {\bibinfo  {journal} {Journal of Statistical Physics}\ }\textbf {\bibinfo {volume} {160}},\ \bibinfo {pages} {1154} (\bibinfo {year} {2015})}\BibitemShut {NoStop}%
\bibitem [{\citenamefont {Gingrich}\ \emph {et~al.}(2016)\citenamefont {Gingrich}, \citenamefont {Horowitz}, \citenamefont {Perunov},\ and\ \citenamefont {England}}]{Gingrich2016}%
  \BibitemOpen
  \bibfield  {author} {\bibinfo {author} {\bibfnamefont {T.~R.}\ \bibnamefont {Gingrich}}, \bibinfo {author} {\bibfnamefont {J.~M.}\ \bibnamefont {Horowitz}}, \bibinfo {author} {\bibfnamefont {N.}~\bibnamefont {Perunov}},\ and\ \bibinfo {author} {\bibfnamefont {J.~L.}\ \bibnamefont {England}},\ }\bibfield  {title} {\bibinfo {title} {Dissipation bounds all steady-state current fluctuations},\ }\href {https://doi.org/10.1103/PhysRevLett.116.120601} {\bibfield  {journal} {\bibinfo  {journal} {Phys. Rev. Lett.}\ }\textbf {\bibinfo {volume} {116}},\ \bibinfo {pages} {120601} (\bibinfo {year} {2016})}\BibitemShut {NoStop}%
\bibitem [{\citenamefont {Pietzonka}\ \emph {et~al.}(2016)\citenamefont {Pietzonka}, \citenamefont {Barato},\ and\ \citenamefont {Seifert}}]{Pietzonka2016}%
  \BibitemOpen
  \bibfield  {author} {\bibinfo {author} {\bibfnamefont {P.}~\bibnamefont {Pietzonka}}, \bibinfo {author} {\bibfnamefont {A.~C.}\ \bibnamefont {Barato}},\ and\ \bibinfo {author} {\bibfnamefont {U.}~\bibnamefont {Seifert}},\ }\bibfield  {title} {\bibinfo {title} {Affinity- and topology-dependent bound on current fluctuations},\ }\href {https://doi.org/10.1088/1751-8113/49/34/34LT01} {\bibfield  {journal} {\bibinfo  {journal} {Journal of Physics A: Mathematical and Theoretical}\ }\textbf {\bibinfo {volume} {49}},\ \bibinfo {pages} {34LT01} (\bibinfo {year} {2016})}\BibitemShut {NoStop}%
\bibitem [{\citenamefont {Polettini}\ \emph {et~al.}(2016)\citenamefont {Polettini}, \citenamefont {Lazarescu},\ and\ \citenamefont {Esposito}}]{Polettini2016}%
  \BibitemOpen
  \bibfield  {author} {\bibinfo {author} {\bibfnamefont {M.}~\bibnamefont {Polettini}}, \bibinfo {author} {\bibfnamefont {A.}~\bibnamefont {Lazarescu}},\ and\ \bibinfo {author} {\bibfnamefont {M.}~\bibnamefont {Esposito}},\ }\bibfield  {title} {\bibinfo {title} {Tightening the uncertainty principle for stochastic currents},\ }\href {https://doi.org/10.1103/PhysRevE.94.052104} {\bibfield  {journal} {\bibinfo  {journal} {Phys. Rev. E}\ }\textbf {\bibinfo {volume} {94}},\ \bibinfo {pages} {052104} (\bibinfo {year} {2016})}\BibitemShut {NoStop}%
\bibitem [{\citenamefont {Gingrich}\ \emph {et~al.}(2017)\citenamefont {Gingrich}, \citenamefont {Rotskoff},\ and\ \citenamefont {Horowitz}}]{Gingrich2017}%
  \BibitemOpen
  \bibfield  {author} {\bibinfo {author} {\bibfnamefont {T.~R.}\ \bibnamefont {Gingrich}}, \bibinfo {author} {\bibfnamefont {G.~M.}\ \bibnamefont {Rotskoff}},\ and\ \bibinfo {author} {\bibfnamefont {J.~M.}\ \bibnamefont {Horowitz}},\ }\bibfield  {title} {\bibinfo {title} {Inferring dissipation from current fluctuations},\ }\href {https://doi.org/10.1088/1751-8121/aa672f} {\bibfield  {journal} {\bibinfo  {journal} {Journal of Physics A: Mathematical and Theoretical}\ }\textbf {\bibinfo {volume} {50}},\ \bibinfo {pages} {184004} (\bibinfo {year} {2017})}\BibitemShut {NoStop}%
\bibitem [{\citenamefont {Gingrich}\ and\ \citenamefont {Horowitz}(2017)}]{GingrichPRL2017}%
  \BibitemOpen
  \bibfield  {author} {\bibinfo {author} {\bibfnamefont {T.~R.}\ \bibnamefont {Gingrich}}\ and\ \bibinfo {author} {\bibfnamefont {J.~M.}\ \bibnamefont {Horowitz}},\ }\bibfield  {title} {\bibinfo {title} {Fundamental bounds on first passage time fluctuations for currents},\ }\href {https://doi.org/10.1103/PhysRevLett.119.170601} {\bibfield  {journal} {\bibinfo  {journal} {Phys. Rev. Lett.}\ }\textbf {\bibinfo {volume} {119}},\ \bibinfo {pages} {170601} (\bibinfo {year} {2017})}\BibitemShut {NoStop}%
\bibitem [{\citenamefont {Garrahan}(2017)}]{Garrahan2017}%
  \BibitemOpen
  \bibfield  {author} {\bibinfo {author} {\bibfnamefont {J.~P.}\ \bibnamefont {Garrahan}},\ }\bibfield  {title} {\bibinfo {title} {Simple bounds on fluctuations and uncertainty relations for first-passage times of counting observables},\ }\href {https://doi.org/10.1103/PhysRevE.95.032134} {\bibfield  {journal} {\bibinfo  {journal} {Phys. Rev. E}\ }\textbf {\bibinfo {volume} {95}},\ \bibinfo {pages} {032134} (\bibinfo {year} {2017})}\BibitemShut {NoStop}%
\bibitem [{\citenamefont {Andrieux}\ and\ \citenamefont {Gaspard}(2007)}]{Andrieux2007}%
  \BibitemOpen
  \bibfield  {author} {\bibinfo {author} {\bibfnamefont {D.}~\bibnamefont {Andrieux}}\ and\ \bibinfo {author} {\bibfnamefont {P.}~\bibnamefont {Gaspard}},\ }\bibfield  {title} {\bibinfo {title} {Fluctuation theorem for currents and schnakenberg network theory},\ }\href {https://doi.org/10.1007/s10955-006-9233-5} {\bibfield  {journal} {\bibinfo  {journal} {Journal of Statistical Physics}\ }\textbf {\bibinfo {volume} {127}},\ \bibinfo {pages} {107} (\bibinfo {year} {2007})}\BibitemShut {NoStop}%
\bibitem [{\citenamefont {Bagrets}\ and\ \citenamefont {Nazarov}(2003)}]{Bagrets2003}%
  \BibitemOpen
  \bibfield  {author} {\bibinfo {author} {\bibfnamefont {D.~A.}\ \bibnamefont {Bagrets}}\ and\ \bibinfo {author} {\bibfnamefont {Y.~V.}\ \bibnamefont {Nazarov}},\ }\bibfield  {title} {\bibinfo {title} {Full counting statistics of charge transfer in coulomb blockade systems},\ }\href {https://doi.org/10.1103/PhysRevB.67.085316} {\bibfield  {journal} {\bibinfo  {journal} {Phys. Rev. B}\ }\textbf {\bibinfo {volume} {67}},\ \bibinfo {pages} {085316} (\bibinfo {year} {2003})}\BibitemShut {NoStop}%
\bibitem [{\citenamefont {Utsumi}(2007)}]{Utsumi2007}%
  \BibitemOpen
  \bibfield  {author} {\bibinfo {author} {\bibfnamefont {Y.}~\bibnamefont {Utsumi}},\ }\bibfield  {title} {\bibinfo {title} {Full counting statistics for the number of electrons in a quantum dot},\ }\href {https://doi.org/10.1103/PhysRevB.75.035333} {\bibfield  {journal} {\bibinfo  {journal} {Phys. Rev. B}\ }\textbf {\bibinfo {volume} {75}},\ \bibinfo {pages} {035333} (\bibinfo {year} {2007})}\BibitemShut {NoStop}%
\bibitem [{\citenamefont {Flindt}\ \emph {et~al.}(2010)\citenamefont {Flindt}, \citenamefont {Novotn\'y}, \citenamefont {Braggio},\ and\ \citenamefont {Jauho}}]{FlindtPRB2010}%
  \BibitemOpen
  \bibfield  {author} {\bibinfo {author} {\bibfnamefont {C.}~\bibnamefont {Flindt}}, \bibinfo {author} {\bibfnamefont {T.~c.~v.}\ \bibnamefont {Novotn\'y}}, \bibinfo {author} {\bibfnamefont {A.}~\bibnamefont {Braggio}},\ and\ \bibinfo {author} {\bibfnamefont {A.-P.}\ \bibnamefont {Jauho}},\ }\bibfield  {title} {\bibinfo {title} {Counting statistics of transport through coulomb blockade nanostructures: High-order cumulants and non-markovian effects},\ }\href {https://doi.org/10.1103/PhysRevB.82.155407} {\bibfield  {journal} {\bibinfo  {journal} {Phys. Rev. B}\ }\textbf {\bibinfo {volume} {82}},\ \bibinfo {pages} {155407} (\bibinfo {year} {2010})}\BibitemShut {NoStop}%
\bibitem [{\citenamefont {Maes}\ and\ \citenamefont {Netočný}(2008)}]{Maes_2008}%
  \BibitemOpen
  \bibfield  {author} {\bibinfo {author} {\bibfnamefont {C.}~\bibnamefont {Maes}}\ and\ \bibinfo {author} {\bibfnamefont {K.}~\bibnamefont {Netočný}},\ }\bibfield  {title} {\bibinfo {title} {Canonical structure of dynamical fluctuations in mesoscopic nonequilibrium steady states},\ }\href {https://doi.org/10.1209/0295-5075/82/30003} {\bibfield  {journal} {\bibinfo  {journal} {Europhysics Letters}\ }\textbf {\bibinfo {volume} {82}},\ \bibinfo {pages} {30003} (\bibinfo {year} {2008})}\BibitemShut {NoStop}%
\bibitem [{Lis()}]{LisSym}%
  \BibitemOpen
  \href@noop {} {}\bibinfo {note} {See Supplemental Material for the list of symbols.}\BibitemShut {Stop}%
\bibitem [{\citenamefont {Hill}(1966)}]{Hill1966}%
  \BibitemOpen
  \bibfield  {author} {\bibinfo {author} {\bibfnamefont {T.~L.}\ \bibnamefont {Hill}},\ }\bibfield  {title} {\bibinfo {title} {Studies in irreversible thermodynamics iv. diagrammatic representation of steady state fluxes for unimolecular systems},\ }\href {https://doi.org/https://doi.org/10.1016/0022-5193(66)90137-8} {\bibfield  {journal} {\bibinfo  {journal} {Journal of Theoretical Biology}\ }\textbf {\bibinfo {volume} {10}},\ \bibinfo {pages} {442} (\bibinfo {year} {1966})}\BibitemShut {NoStop}%
\bibitem [{\citenamefont {Schnakenberg}(1976)}]{Schnakenberg1976}%
  \BibitemOpen
  \bibfield  {author} {\bibinfo {author} {\bibfnamefont {J.}~\bibnamefont {Schnakenberg}},\ }\bibfield  {title} {\bibinfo {title} {Network theory of microscopic and macroscopic behavior of master equation systems},\ }\href {https://doi.org/10.1103/RevModPhys.48.571} {\bibfield  {journal} {\bibinfo  {journal} {Re1978v. Mod. Phys.}\ }\textbf {\bibinfo {volume} {48}},\ \bibinfo {pages} {571} (\bibinfo {year} {1976})}\BibitemShut {NoStop}%
\bibitem [{\citenamefont {Weidlich}(1978)}]{Weidlich1978}%
  \BibitemOpen
  \bibfield  {author} {\bibinfo {author} {\bibfnamefont {W.}~\bibnamefont {Weidlich}},\ }\bibfield  {title} {\bibinfo {title} {On the structure of exact solutions of discrete masterequations},\ }\href {https://doi.org/10.1007/BF01320040} {\bibfield  {journal} {\bibinfo  {journal} {Zeitschrift für Physik B Condensed Matter}\ }\textbf {\bibinfo {volume} {30}},\ \bibinfo {pages} {345} (\bibinfo {year} {1978})}\BibitemShut {NoStop}%
\bibitem [{\citenamefont {Pal}\ \emph {et~al.}(2021{\natexlab{a}})\citenamefont {Pal}, \citenamefont {Reuveni},\ and\ \citenamefont {Rahav}}]{Pal2021}%
  \BibitemOpen
  \bibfield  {author} {\bibinfo {author} {\bibfnamefont {A.}~\bibnamefont {Pal}}, \bibinfo {author} {\bibfnamefont {S.}~\bibnamefont {Reuveni}},\ and\ \bibinfo {author} {\bibfnamefont {S.}~\bibnamefont {Rahav}},\ }\bibfield  {title} {\bibinfo {title} {Thermodynamic uncertainty relation for systems with unidirectional transitions},\ }\href {https://doi.org/10.1103/PhysRevResearch.3.013273} {\bibfield  {journal} {\bibinfo  {journal} {Phys. Rev. Res.}\ }\textbf {\bibinfo {volume} {3}},\ \bibinfo {pages} {013273} (\bibinfo {year} {2021}{\natexlab{a}})}\BibitemShut {NoStop}%
\bibitem [{\citenamefont {Shiraishi}(2021)}]{Shiraishi2021}%
  \BibitemOpen
  \bibfield  {author} {\bibinfo {author} {\bibfnamefont {N.}~\bibnamefont {Shiraishi}},\ }\bibfield  {title} {\bibinfo {title} {Optimal thermodynamic uncertainty relation in markov jump processes},\ }\href {https://doi.org/10.1007/s10955-021-02829-8} {\bibfield  {journal} {\bibinfo  {journal} {Journal of Statistical Physics}\ }\textbf {\bibinfo {volume} {185}},\ \bibinfo {pages} {19} (\bibinfo {year} {2021})}\BibitemShut {NoStop}%
\bibitem [{\citenamefont {Barato}\ \emph {et~al.}(2018)\citenamefont {Barato}, \citenamefont {Chetrite}, \citenamefont {Faggionato},\ and\ \citenamefont {Gabrielli}}]{Barato2018}%
  \BibitemOpen
  \bibfield  {author} {\bibinfo {author} {\bibfnamefont {A.~C.}\ \bibnamefont {Barato}}, \bibinfo {author} {\bibfnamefont {R.}~\bibnamefont {Chetrite}}, \bibinfo {author} {\bibfnamefont {A.}~\bibnamefont {Faggionato}},\ and\ \bibinfo {author} {\bibfnamefont {D.}~\bibnamefont {Gabrielli}},\ }\bibfield  {title} {\bibinfo {title} {Bounds on current fluctuations in periodically driven systems},\ }\href {https://doi.org/10.1088/1367-2630/aae512} {\bibfield  {journal} {\bibinfo  {journal} {New Journal of Physics}\ }\textbf {\bibinfo {volume} {20}},\ \bibinfo {pages} {103023} (\bibinfo {year} {2018})}\BibitemShut {NoStop}%
\bibitem [{\citenamefont {Gross}\ and\ \citenamefont {Kotiuga}(2004)}]{GrossBook2004}%
  \BibitemOpen
  \bibfield  {author} {\bibinfo {author} {\bibfnamefont {P.~W.}\ \bibnamefont {Gross}}\ and\ \bibinfo {author} {\bibfnamefont {P.~R.}\ \bibnamefont {Kotiuga}},\ }\href@noop {} {\emph {\bibinfo {title} {Electromagnetic theory and computation: a topological approach}}},\ \bibinfo {series} {Mathematical Sciences Research Institute Publications}, Vol.~\bibinfo {volume} {48}\ (\bibinfo  {publisher} {Cambridge University Press},\ \bibinfo {address} {Cambridge},\ \bibinfo {year} {2004})\BibitemShut {NoStop}%
\bibitem [{\citenamefont {Nagaosa}(1999)}]{Nagaosa1999}%
  \BibitemOpen
  \bibfield  {author} {\bibinfo {author} {\bibfnamefont {N.}~\bibnamefont {Nagaosa}},\ }\bibinfo {title} {Problems related to superconductivity},\ in\ \href {https://doi.org/10.1007/978-3-662-03774-4_5} {\emph {\bibinfo {booktitle} {Quantum Field Theory in Condensed Matter Physics}}}\ (\bibinfo  {publisher} {Springer Berlin Heidelberg},\ \bibinfo {address} {Berlin, Heidelberg},\ \bibinfo {year} {1999})\ pp.\ \bibinfo {pages} {113--160}\BibitemShut {NoStop}%
\bibitem [{K5()}]{K5}%
  \BibitemOpen
  \href@noop {} {}\bibinfo {note} {The discussion applies to both planar and non-planar graphs. For an example, see the Supplemental Material.}\BibitemShut {Stop}%
\bibitem [{\citenamefont {Pal}\ \emph {et~al.}(2021{\natexlab{b}})\citenamefont {Pal}, \citenamefont {Reuveni},\ and\ \citenamefont {Rahav}}]{PalPRR2021}%
  \BibitemOpen
  \bibfield  {author} {\bibinfo {author} {\bibfnamefont {A.}~\bibnamefont {Pal}}, \bibinfo {author} {\bibfnamefont {S.}~\bibnamefont {Reuveni}},\ and\ \bibinfo {author} {\bibfnamefont {S.}~\bibnamefont {Rahav}},\ }\bibfield  {title} {\bibinfo {title} {Thermodynamic uncertainty relation for first-passage times on markov chains},\ }\href {https://doi.org/10.1103/PhysRevResearch.3.L032034} {\bibfield  {journal} {\bibinfo  {journal} {Phys. Rev. Res.}\ }\textbf {\bibinfo {volume} {3}},\ \bibinfo {pages} {L032034} (\bibinfo {year} {2021}{\natexlab{b}})}\BibitemShut {NoStop}%
\bibitem [{\citenamefont {Cormen}\ \emph {et~al.}(2022)\citenamefont {Cormen}, \citenamefont {Leiserson}, \citenamefont {Rivest},\ and\ \citenamefont {Stein}}]{Cormen_Thomas_H_2022-04-05}%
  \BibitemOpen
  \bibfield  {author} {\bibinfo {author} {\bibfnamefont {T.~H.}\ \bibnamefont {Cormen}}, \bibinfo {author} {\bibfnamefont {C.~E.}\ \bibnamefont {Leiserson}}, \bibinfo {author} {\bibfnamefont {R.~L.}\ \bibnamefont {Rivest}},\ and\ \bibinfo {author} {\bibfnamefont {C.}~\bibnamefont {Stein}},\ }\href {https://lead.to/amazon/jp/?op=bt&la=ja&key=026204630X} {\emph {\bibinfo {title} {Introduction to Algorithms, fourth edition}}},\ \bibinfo {edition} {fourth edition}\ ed.\ (\bibinfo  {publisher} {The MIT Press},\ \bibinfo {year} {2022})\ p.\ \bibinfo {pages} {1312}\BibitemShut {NoStop}%
\end{thebibliography}%


\begin{thebibliography}{1}%
\makeatletter
\providecommand \@ifxundefined [1]{%
 \@ifx{#1\undefined}
}%
\providecommand \@ifnum [1]{%
 \ifnum #1\expandafter \@firstoftwo
 \else \expandafter \@secondoftwo
 \fi
}%
\providecommand \@ifx [1]{%
 \ifx #1\expandafter \@firstoftwo
 \else \expandafter \@secondoftwo
 \fi
}%
\providecommand \natexlab [1]{#1}%
\providecommand \enquote  [1]{``#1''}%
\providecommand \bibnamefont  [1]{#1}%
\providecommand \bibfnamefont [1]{#1}%
\providecommand \citenamefont [1]{#1}%
\providecommand \href@noop [0]{\@secondoftwo}%
\providecommand \href [0]{\begingroup \@sanitize@url \@href}%
\providecommand \@href[1]{\@@startlink{#1}\@@href}%
\providecommand \@@href[1]{\endgroup#1\@@endlink}%
\providecommand \@sanitize@url [0]{\catcode `\\12\catcode `\$12\catcode `\&12\catcode `\#12\catcode `\^12\catcode `\_12\catcode `\%12\relax}%
\providecommand \@@startlink[1]{}%
\providecommand \@@endlink[0]{}%
\providecommand \url  [0]{\begingroup\@sanitize@url \@url }%
\providecommand \@url [1]{\endgroup\@href {#1}{\urlprefix }}%
\providecommand \urlprefix  [0]{URL }%
\providecommand \Eprint [0]{\href }%
\providecommand \doibase [0]{https://doi.org/}%
\providecommand \selectlanguage [0]{\@gobble}%
\providecommand \bibinfo  [0]{\@secondoftwo}%
\providecommand \bibfield  [0]{\@secondoftwo}%
\providecommand \translation [1]{[#1]}%
\providecommand \BibitemOpen [0]{}%
\providecommand \bibitemStop [0]{}%
\providecommand \bibitemNoStop [0]{.\EOS\space}%
\providecommand \EOS [0]{\spacefactor3000\relax}%
\providecommand \BibitemShut  [1]{\csname bibitem#1\endcsname}%
\let\auto@bib@innerbib\@empty
\bibitem [{\citenamefont {Flindt}\ \emph {et~al.}(2010)\citenamefont {Flindt}, \citenamefont {Novotn\'y}, \citenamefont {Braggio},\ and\ \citenamefont {Jauho}}]{FlindtPRB2010s}%
  \BibitemOpen
  \bibfield  {author} {\bibinfo {author} {\bibfnamefont {C.}~\bibnamefont {Flindt}}, \bibinfo {author} {\bibfnamefont {T.~c.~v.}\ \bibnamefont {Novotn\'y}}, \bibinfo {author} {\bibfnamefont {A.}~\bibnamefont {Braggio}},\ and\ \bibinfo {author} {\bibfnamefont {A.-P.}\ \bibnamefont {Jauho}},\ }\bibfield  {title} {\bibinfo {title} {Counting statistics of transport through coulomb blockade nanostructures: High-order cumulants and non-markovian effects},\ }\href {https://doi.org/10.1103/PhysRevB.82.155407} {\bibfield  {journal} {\bibinfo  {journal} {Phys. Rev. B}\ }\textbf {\bibinfo {volume} {82}},\ \bibinfo {pages} {155407} (\bibinfo {year} {2010})}\BibitemShut {NoStop}%
\end{thebibliography}


\clearpage

\renewcommand{\appendixname}{}




\begin{titlepage}
\title{Supplemental Material for \\ `Network analysis for the steady-state thermodynamic uncertainty relation'}
\maketitle
\end{titlepage}
\onecolumngrid


List of Symbols and Example: 5-Clique-Like Graph.


\renewcommand{\thesection}{S\arabic{section}}
\setcounter{figure}{0}
\setcounter{section}{0}

\section{List of Symbols}

In the following, the symbol $\sqcup$ represents the union of mutually disjoint sets. 


\begin{tabularx}{\textwidth}{XX}
$G_{\rm m}=(V,E_{\rm m})$ & Directed multigraph \\
$V$ & Set of nodes \\
$E_{\rm m} = E_{\rm uni} \sqcup E_{\rm bi}$ & Set of arcs (directed edges) \\
$e=(v_- \leftarrow v_+)$ & Arc from the tail node $v_+ \in V$ to the head node $v_- \in V$ \\
$\partial^\pm$ & Boundary operator: $\partial^\pm e = \partial^\pm (v_- \leftarrow v_+) = v_\pm$ \\
$D^{\pm}_{v,e} = \delta_{v,\partial^\pm e}$ & $v$, $e$ component of positive/negative incidence matrix \\
$D=D^{+}-D^{-}$ & Incidence matrix \\
$\Gamma_e > 0$ & Transition rate associated to arc $e \in E_{\rm m}$ \\
$W_{e}$ & Number of transitions through an arc $e$ in its direction \\
$\tau_v$ & Sojourn time at node $v \in V$ \\
$\lambda_{e}=-i x_e$ & Counting field for $W_e$ \\
$\xi_v=-i y_v$ & Counting field for $\tau_v$ \\
$w_{e} = W_{e}/\tau$ & (Empirical) flux \\
$n_{v} = \tau_{v}/\tau$ & (Empirical) node state probability (The symbol $n_v$ is sometimes used to denote the average, depending on the context). \\
$j_e = w_e - w_{-e}$ & Anti-symmetrized flux (empirical edge current) \\
$g_e = w_e + w_{-e}$ & Symmetrized flux \\
$a(n_v)=\Gamma_e n_{\partial^+e}$ & Activity rate \\
$j_e(n_v) = a_e(n_v) - a_{-e}(n_v)$ & Edge current \\ 
$g_e(n_v) = a_e(n_v)+a_{-e}(n_v)$ & Traffic, which is the reciprocal of the weight of edge $e$\\
$n_v^{\rm st}$ & Node state probability in the steady-state \\
${\mathcal P}_\tau ( \{ W_{e} \} , \{ \tau_v \} )$ & Joint probability distribution during the measurement time $\tau$ \\
${\mathcal I}$ & Large deviation function (rate function) \\
\end{tabularx}

\begin{tabularx}{\textwidth}{XX}
$\overline{A} = \left \{ -e \middle | e \in A \right \}$ & Set of reversed arcs of a set $A$ \\
$E_{\rm uni} = \left \{ e \middle | e \in E_{\rm m} \wedge -e \notin E_{\rm m}  \right \}$ & Set of arcs for unidirectional transitions \\
$E_{\rm bi} = \left \{ e \middle | e \in E_{\rm m}  \wedge -e \in E_{\rm m} \right \} = E_{\rm b} \sqcup \overline{E_{\rm b}}$  & Set of arcs for bidirectional transitions \\
$E_{\rm b}$ & Set of arcs for forward transitions among bidirectional transitions \\
$G=(V,E= E_{\rm uni} \sqcup E_{\rm b})$ & Oriented graph \\
$E= E_{\rm uni} \sqcup E_{\rm b}$ & Set of arcs of the oriented graph $G$ \\
$T=(V(T)=V,E(T))$ & Directed rooted spanning tree \\
$V(T)$ & Set of nodes of tree $T$ \\ 
$E(T)$ & Set of arcs of tree $T$ \\ 
$T^*$ & Cotree of $T$ \\
\end{tabularx}

\begin{tabularx}{\textwidth}{XX}
$\mathbb{R}$ & Set of real numbers \\
$\mathbb{R}^M$ & Set of $M$ component real vectors \\
$\mathbb{R}^{M \times N}$ & Set of $M \times N$ real matrices \\
$|A|$ & Number of elements of the set $A$ \\
\end{tabularx}

\begin{tabularx}{\textwidth}{XX}
${\bm j} = \begin{bmatrix} {\bm j}_t \\ {\bm j}_{c} \end{bmatrix} \in \mathbb{R}^{|E|}$ & $|E|=|E(T)|+|E(T^*)|$ component edge current vector \\
${{\bm j}_t}^T = \begin{bmatrix} j_{t_1}, \cdots, j_{t_{|E(T)|}} \end{bmatrix}$, ($t_1,\cdots, t_{|E(T)|} \in E(T)$) & \\
${{\bm j}_c}^T = \begin{bmatrix} j_{c_1}, \cdots, j_{c_{|E(T^*)|}} \end{bmatrix}$, ($c_1,\cdots, c_{|E(T^*)|} \in E(T^*)$) & \\
${\bm n}^T = \begin{bmatrix} n_{v_1}, \cdots, n_{v_{|V|}} \end{bmatrix}$, ($v_1, \cdots, v_{|V|} \in V$) & State probability vector \\
${\bm G} ={\rm diag} \,{\bm g}({\bm n}^{\rm st})$, (${\bm g}({\bm n}^{\rm st}) \in \mathbb{R}^{|E|}$) & Inverse of weight matrix \\
\end{tabularx}

\begin{tabularx}{\textwidth}{XX}
${\bm 1}$ & Real vector with all entries equal to one \\
${\bm e}_e \in \mathbb{R}^{|E|}$ & Unit vector: $\left( {\bm e}_e \right)_{e'} = \delta_{e,e'}$ \\
\end{tabularx}

\begin{tabularx}{\textwidth}{XX}
${\bm I}_{M}$ & $M \times M$ unit matrix \\
${\bm 0}_{M \times N}$ & $M \times N$ zero matrix \\
$\mathbbm{1}_{A}(a)$ & Indicator function, which equals $1$ if $a \in A$ and equals $0$ if $a \notin A$. \\
$C_{e_1} =(e_{\ell}, \cdots,e_3,e_2,e_1)$ & Fundamental cycle of length $\ell$, which is a sequence of one chord $e_1 \in E(T^*)$ and $\ell-1$ twigs $e_{2}, e_{3}, \cdots, e_{\ell} \in E(T)$ or reversed ones $e_{2}, e_{3}, \cdots, e_{\ell}  \in \overline{E(T)}$. The head and tail of adjacent arcs share the same node $\partial^- e_{n} = \partial^+ e_{n+1}$ ($n=1,\cdots,\ell-1$). The first and last arcs satisfy the periodic boundary condition $\partial^- e_\ell= \partial^+ e_1$. \\
$P_{v \leftarrow v_0} =(e_\ell, \cdots , e_2,e_1)$ & Directed path from the root $v_0$ to a node $v$ along the directed rooted spanning tree $T$ ($e_1,e_2,\cdots e_\ell \in E(T)$). The head and tail of adjacent arcs share the same node $\partial^- e_{n} = \partial^+ e_{n+1}$ ($n=1,\cdots,\ell-1$). The two endpoints are $\partial^+ e_1=v_0$ and $\partial^- e_\ell=v$. \\
${\bm B} = \begin{bmatrix} -{\bm F}^T & {\bm I}_{|E(T^*)|} \end{bmatrix}  \in \mathbb{R}^{|E(T^*)| \times |E|}$ & Fundamental cycle matrix: $\left( {\bm F} \right)_{t,c} = \mathbbm{1}_{C_c}(-t) - \mathbbm{1}_{C_c}(t)$ for $c \in E(T^*)$ and $t \in E(T)$. \\
${\bm S} \in \mathbb{R}^{|V| \times |E(T)|}$ & Root-to-node path matrix: $\left( {\bm S} \right)_{v,t} = \mathbbm{1}_{P_{v \leftarrow v_0}}(t)$ for $v \in V$ and $t \in E(T)$. \\
${\bm Q} \in \mathbb{R}^{|E(T)| \times |E|}$ & Fundamental cutset matrix: $-{\bm S}^T {\bm D}  = {\bm Q} =  \begin{bmatrix} {\bm I}_{|E(T)|} & {\bm F} \end{bmatrix}$. \\ 
\end{tabularx}

\begin{tabularx}{\textwidth}{XX}
$T_{\partial^- t}$ & Subtree rooted at $\partial^- t$ obtained by cutting the directed rooted spanning tree $T$ by removing the arc $t$ \\
\end{tabularx}

\begin{tabularx}{\textwidth}{XX}
$\bigcup_{d=0}^\ell A_d = A_0 \cup A_1 \cup \cdots \cup A_\ell$ & \\
\end{tabularx}

\renewcommand{\theequation}{S\arabic{equation}}
\renewcommand{\thefigure}{S\arabic{figure}}

\section{Example: 5-Clique-Like Graph}

\begin{figure}[ht]
\begin{center}
\includegraphics[width=0.3 \columnwidth]{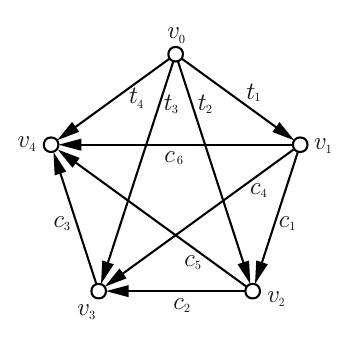}
\caption{
A 5-clique-like Graph.
}
\label{fig:K5}
\end{center}
\end{figure}

The 5-clique is a non-planer graph consisting of 5 nodes, $V=\{ v_0, v_1, v_2, v_3, v_4 \}$, 
and 10 edges connecting all possible pairs of nodes. 
We assign a direction to each edge as illustrated in Fig.~\ref{fig:K5} and introduce a directed rooted spanning tree $T$ rooted at $v_0$ formed with the set of twigs $E(T)=\{ t_1,t_2,t_3,t_4 \}$. 
The set of chords is $E(T^*)=\{ c_1, c_2, c_3, c_4, c_5, c_{6} \}$. 
We partition the set of arcs into a singleton set of a unidirectional transition, $E_{\rm uni}=\{ c_3 \}$ and the set of bidirectional transitions, $E_{\rm b}=E(T) \cup E(T^*) -E_{\rm uni}$. 
We measure the current flowing through the chord $c_3$: 
\begin{align}
{\bm d}^T
=
\begin{blockarray}{*{11}{c}}
& \scriptstyle t_1 & \scriptstyle t_2 & \scriptstyle t_3 & \scriptstyle t_4 & \scriptstyle c_1 & \scriptstyle c_2 & \scriptstyle c_3 & \scriptstyle c_4 & \scriptstyle c_5 & \scriptstyle c_6 \\
\begin{block}{c[cccccccccc]}
&   0 &   0 &   0 &   0 &   0 &   0 &   1 &   0 &   0 &   0  \\
\end{block}
\end{blockarray}
\, .
\end{align}
The incidence matrix ${\bm D}$, the matrix ${\bm F}$, and the root-to-node path matrix ${\bm S}$ are, 
\begin{align}
{\bm D}
=
\begin{blockarray}{*{11}{c}}
& \scriptstyle t_1 & \scriptstyle t_2 & \scriptstyle t_3 & \scriptstyle t_4 & \scriptstyle c_1 & \scriptstyle c_2 & \scriptstyle c_3 & \scriptstyle c_4 & \scriptstyle c_5 & \scriptstyle c_6 \\
\begin{block}{c[cccccccccc]}
\bigstrutht {\scriptstyle v_0} &   1 &   1 &   1 &   1 &   0 &   0 &   0 &   0 &   0 &   0    \\
{\scriptstyle v_1} &  -1 &   0 &   0 &   0 &   1 &   0 &   0 &   1 &   0 &   1    \\
{\scriptstyle v_2} &   0 &  -1 &   0 &   0 &  -1 &   1 &   0 &   0 &   1 &   0    \\
{\scriptstyle v_3} &   0 &   0 &  -1 &   0 &   0 &  -1 &   1 &  -1 &   0 &   0    \\
\bigstrutdp {\scriptstyle v_4} &   0 &   0 &   0 &  -1 &   0 &   0 &  -1 &   0 &  -1 &  -1    \\
\end{block}
\end{blockarray}
\, , 
\;\;\;\;
-{\bm F}^T
=
\begin{blockarray}{*{5}{c}}
& \scriptstyle t_1 & \scriptstyle t_2 & \scriptstyle t_3 & \scriptstyle t_4 \\
\begin{block}{c[cccc]}
\bigstrutht \scriptstyle C_{c_1}    &   1 &  -1 &   0 &   0 \\
\scriptstyle C_{c_2}    &   0 &   1 &  -1 &   0 \\
\scriptstyle C_{c_3}    &   0 &   0 &   1 &  -1 \\
\scriptstyle C_{c_4}    &   1 &   0 &  -1 &   0 \\
\scriptstyle C_{c_5}    &   0 &   1 &   0 &  -1 \\
\bigstrutdp \scriptstyle C_{c_6}    &   1 &   0 &   0 &  -1 \\
\end{block}
\end{blockarray}
\, , 
\;\;\;\;
{\bm S}
=
\begin{blockarray}{*{5}{c}}
& \scriptstyle t_1 & \scriptstyle t_2 & \scriptstyle t_3 & \scriptstyle t_4 \\
\begin{block}{c[cccc]}
\bigstrutht 
\scriptstyle v_0 &   0 &   0 &   0 &   0  \\
\scriptstyle v_1 &   1 &   0 &   0 &   0  \\
\scriptstyle v_2 &   0 &   1 &   0 &   0  \\
\scriptstyle v_3 &   0 &   0 &   1 &   0  \\
\bigstrutdp \scriptstyle v_4 &   0 &   0 &   0 &   1  \\
\end{block}
\end{blockarray}
\, .
\end{align}
The transition rate matrix is, 
\begin{align}
{\bm L} = \sum_\pm 
\left (
{\bm D}^\mp {\bm \Pi}_{E_{\rm b}} { {\bm D}^\pm }^T
-
{\bm D}^\pm {\bm \Pi}_{E_{\rm b}} { {\bm D}^\pm }^T
\right )
+
{\bm D}^- {\bm \Pi}_{E_{\rm uni}} {{\bm D}^+}^T
-
{\bm D}^+ {\bm \Pi}_{E_{\rm uni}} {{\bm D}^+}^T
\, ,
\label{eqn:LK5}
\end{align}
where we write the projection operator as, 
${\bm \Pi}_{A} = \sum_{e \in A} {\bm e}_e^T {\bm e}_e$. 
Then, the state probability vector in the steady state becomes, 
\begin{align}
{ {\bm n}^{\rm st} }^T = 
\begin{blockarray}{*{6}{c}}
& \scriptstyle v_0 & \scriptstyle v_1 & \scriptstyle v_2 & \scriptstyle v_3 & \scriptstyle v_4 \\
\begin{block}{c[ccccc]}
& \displaystyle \frac{1}{5} & \displaystyle \frac{1}{5} & \displaystyle \frac{1}{5}  &  \displaystyle \frac{3}{20}  &  \displaystyle \frac{1}{4}  \\
\end{block}
\end{blockarray}
\, , 
\end{align}
which results in the average current $\langle \! \langle w \rangle \! \rangle =3/20$. 
An optimal point ${\bm f}^*$ minimizing the expression inside the infimum in (\ref{eqn:sig_2chain}) under the constraint $\langle \! \langle w \rangle \! \rangle = \left( {\bm f} \right)_{C_{c_3}}$ is
\begin{align}
{{\bm f}^*}^T = 
\begin{blockarray}{*{7}{c}}
& \scriptstyle C_{c_1} & \scriptstyle C_{c_2} & \scriptstyle C_{c_3} & \scriptstyle C_{c_4} & \scriptstyle C_{c_5} & \scriptstyle C_{c_6} \\
\begin{block}{c[cccccc]}
& 0 & \displaystyle \frac{1}{20} & \langle \! \langle w \rangle \! \rangle & \displaystyle \frac{1}{20} & \displaystyle -\frac{1}{20} & \displaystyle -\frac{1}{20} \\
\end{block}
\end{blockarray}
\, .
\end{align}
Then the right-hand-side of (\ref{eqn:sig_2chain}) becomes,  
\begin{align}
\inf_{ {\bm f} \in F(w)} \left( {\bm f}^T {\bm B} {\bm \Phi}^T {\bm B}^T {\bm G}_{2}^{-1} {\bm B} {\bm \Phi} {\bm B}^T {\bm f} \right)
=&
\langle \! \langle w \rangle \! \rangle^2
\left[
{\bm P}_{C_{c_3}} {\bm M} {\bm P}_{C_{c_3}}^T
-
\left({\bm P}_{C_{c_3}} {\bm M} {\bm Q}_{C_{c_3}}^T \right) 
\left({\bm Q}_{C_{c_3}} {\bm M} {\bm Q}_{C_{c_3}}^T \right)^{-1}
\left({\bm Q}_{C_{c_3}} {\bm M} {\bm P}_{C_{c_3}}^T \right) 
\right]
\nonumber \\
&= \frac{30}{173} \approx 0.173 \, , \label{eqn:invFano}
\end{align}
where
${\bm M} =  {\bm B} {\bm \Phi}^T {\bm B}^T {\bm G}_2^{-1} {\bm B} {\bm \Phi} {\bm B}^T$
and 
\begin{align}
{\bm P}_{C_{c_3}} =
\begin{blockarray}{*{7}{c}}
& \scriptstyle C_{c_1} & \scriptstyle C_{c_2} & \scriptstyle C_{c_3} & \scriptstyle C_{c_4} & \scriptstyle C_{c_5} & \scriptstyle C_{c_6} \\
\begin{block}{c[cccccc]}
\scriptstyle C_{c_3} & 0 & 0 & 1 & 0 & 0 & 0  \\
\end{block}
\end{blockarray}
\, ,
\;\;\;\;
{\bm Q}_{C_{c_3}}
=
\begin{blockarray}{*{7}{c}}
& \scriptstyle C_{c_1} & \scriptstyle C_{c_2} & \scriptstyle C_{c_3} & \scriptstyle C_{c_4} & \scriptstyle C_{c_5} & \scriptstyle C_{c_6} \\
\begin{block}{c[cccccc]}
\bigstrutht 
\scriptstyle C_{c_1} & 1 & 0 & 0 & 0 & 0 & 0  \\
\scriptstyle C_{c_2} & 0 & 1 & 0 & 0 & 0 & 0  \\
\scriptstyle C_{c_4} & 0 & 0 & 0 & 1 & 0 & 0  \\
\scriptstyle C_{c_5} & 0 & 0 & 0 & 0 & 1 & 0  \\
\bigstrutdp 
\scriptstyle C_{c_6} & 0 & 0 & 0 & 0 & 0 & 1  \\
\end{block}
\end{blockarray}
\, .
\end{align}

A deviation from this optimal point results in a larger value. 
For example, omitting `internal' cycle currents, 
\begin{align}
{{\bm f}_{0}}^T = 
\begin{blockarray}{*{7}{c}}
& \scriptstyle C_{c_1} & \scriptstyle C_{c_2} & \scriptstyle C_{c_3} & \scriptstyle C_{c_4} & \scriptstyle C_{c_5} & \scriptstyle C_{c_6} \\
\begin{block}{c[cccccc]}
& 0 & 0 & \displaystyle \langle \! \langle w \rangle \! \rangle & 0 & 0 & 0 \\
\end{block}
\end{blockarray}
\, ,
\end{align}
results in,  
\begin{align}
{{\bm f}_{0}}^T {\bm B} {\bm \Phi}^T {\bm B}^T {\bm G}_2^{-1} {\bm B} {\bm \Phi} {\bm B}^T {{\bm f}_{0}} 
=
\langle \! \langle w \rangle \! \rangle^2 {\bm P}_{C_{c_3}} {\bm M} {\bm P}_{C_{c_3}}^T 
= 
\frac{13662}{54841} \approx 0.249 \, ,
\end{align}
which is larger than Eq.~(\ref{eqn:invFano}).

To cross-check the analytic expression~(\ref{eqn:invFano}), we calculate the first and second cumulants using the recursive method~\cite{FlindtPRB2010s}, which is the Rayleigh-Schr\"odinger perturbation theory in terms of the counting field. 
In the present case, the counting field for the current through arc $c_3$, denoted by $\lambda_{c_3}$, is incorporated into the unidirectional transition rate as $e^{i \lambda_{c_3}} {\bm D}^- {\bm \Pi}_{E_{\rm uni}} {{\bm D}^+}^T$ in Eq.~(\ref{eqn:LK5}). 
Then, the first and second cumulants are, 
\begin{align}
\langle \! \langle w \rangle \! \rangle ={\bm 1}^T {\bm L}^{(1)} {\bm n}^{\rm st} = \frac{3}{20} \, ,
\;\;\;\;
\langle \! \langle w^2 \rangle \! \rangle ={\bm 1}^T \left( {\bm L}^{(2)} -2 {\bm L}^{(1)} {\bm R} {\bm L}^{(1)} \right) {\bm n}^{\rm st} = \frac{519}{4000} \, ,
\label{eqn:c1_c2}
\end{align}
where, 
\begin{align}
{\bm L}^{(1)} = {\bm L}^{(2)} = {\bm D}^- {\bm \Pi}_{E_{\rm uni}} {{\bm D}^+}^T
\, .
\end{align}
The pseudo-inverse is 
\begin{align}
{\bm R}= \sum_{\Lambda \neq 0} \frac{ {\bm u}^R_\Lambda { {\bm u}^L_\Lambda }^T }{\Lambda} \, ,
\;\;\;\;
{ {\bm u}^L_\Lambda }^T  {\bm u}^R_{\Lambda'} = \delta_{\Lambda, \Lambda'}
\, ,
\end{align}
where ${\bm u}^{L(R)}_\Lambda$ denotes the left(right) eigenvector of ${\bm L}$ associated with the eigenvalue $\Lambda$. 
They are orthonormalized using the Gram-Schmidt process.
The zero eigenvalue $\Lambda=0$ is excluded from the summation. 
The left and right eigenvectors corresponding to zero eigenvalue are ${\bm u}^{L}_0={\bm 1}$ and ${\bm u}^{R}_0={\bm n}^{\rm st}$, respectively. 
The result (\ref{eqn:c1_c2}) reproduces Eq.~(\ref{eqn:invFano}), 
$\langle \! \langle w \rangle \! \rangle^2 / \langle \! \langle w^2 \rangle \! \rangle=30/173$.

%

\end{document}